\documentclass[useAMS,usenatbib,usegraphicx]{mn2e}
\usepackage{astroshortcuts} 
\usepackage{amssymb} 
\author[M. Zemp et al.]{Marcel Zemp$^1$\thanks{mzemp@ucolick.org}, Ben Moore$^2$, Joachim Stadel$^2$, C. Marcella Carollo$^3$ \& Piero Madau$^1$ \\
$^1$ Department of Astronomy \& Astrophysics, University of California Santa Cruz, 1156 High Street, 95064 Santa Cruz, USA\\
$^2$ Institute for Theoretical Physics, University Zurich, Winterthurerstrasse 190, 8057 Zurich, Switzerland\\
$^3$ Institute of Astronomy, ETH Zurich, Wolfgang-Pauli-Strasse 16, 8093 Zurich, Switzerland}

\title{Multimass spherical structure models for $N$-body simulations}

\begin{document}

\pagerange{\pageref{firstpage}--\pageref{lastpage}} \pubyear{2008}

\maketitle

\label{firstpage}

\begin{abstract}
We present a simple and efficient method to set up spherical structure models for $N$-body simulations with a multimass technique. This technique reduces by a substantial factor the computer run time needed in order to resolve a given scale as compared to single-mass models. It therefore allows to resolve smaller scales in $N$-body simulations for a given computer run time. Here, we present several models with an effective resolution of up to $1.68 \times 10^9$ particles within their virial radius which are stable over cosmologically relevant time-scales. As an application, we confirm the theoretical prediction by \citet{2005MNRAS.360..892D} that in mergers of collisonless structures like dark matter haloes always the cusp of the steepest progenitor is preserved. We model each merger progenitor with an effective number of particles of approximately $10^{8}$ particles. We also find that in a core-core merger the central density approximately doubles whereas in the cusp-cusp case the central density only increases by approximately 50\%. This may suggest that the central region of flat structures are better protected and get less energy input through the merger process.
\end{abstract}


\section{Introduction and Motivation}

Resolution is a key issue in $N$-body simulations. In state-of-the-art cosmological $N$-body simulations today, structures can be fully resolved down to the scale of a fraction of a per cent of the virial radius \citep{2007ApJ...657..262D}. But there are many problems where even higher resolution in central regions of structures is needed. Also, one often would like to study a certain problem without the cosmological context, i.e. one needs a possibility to set up isolated structures with high resolution in order to study dynamical effects in detail and with precise control of the initial condition, which can be very difficult within the framework of a cosmological $N$-body simulation.

For example, the question of whether the central dark matter density profile of haloes that form in cosmological $N$-body simulations is cuspy or cored needs at least a resolution of $\approx 10^{-3}~\rvir$ to be answered. Another example are centrally flat profiles: in order to resolve isolated structures with flat central profiles, a lot of particles are needed since in flat profiles the resolution scaling with the number of particles is the slowest (see below for more details about the scaling of resolution with the number of particles). Another problem we had in mind when developing the multimass technique presented in this paper, was the dynamics of super-massive black hole binaries in the centre of remnants of galaxy mergers. There, the relevant scales are of order of a few pc $\approx 10^{-6}-10^{-5}~\rvir$ for Milky Way size dark matter haloes.

\begin{figure*}
	\centering
	\includegraphics[width=0.49\textwidth]{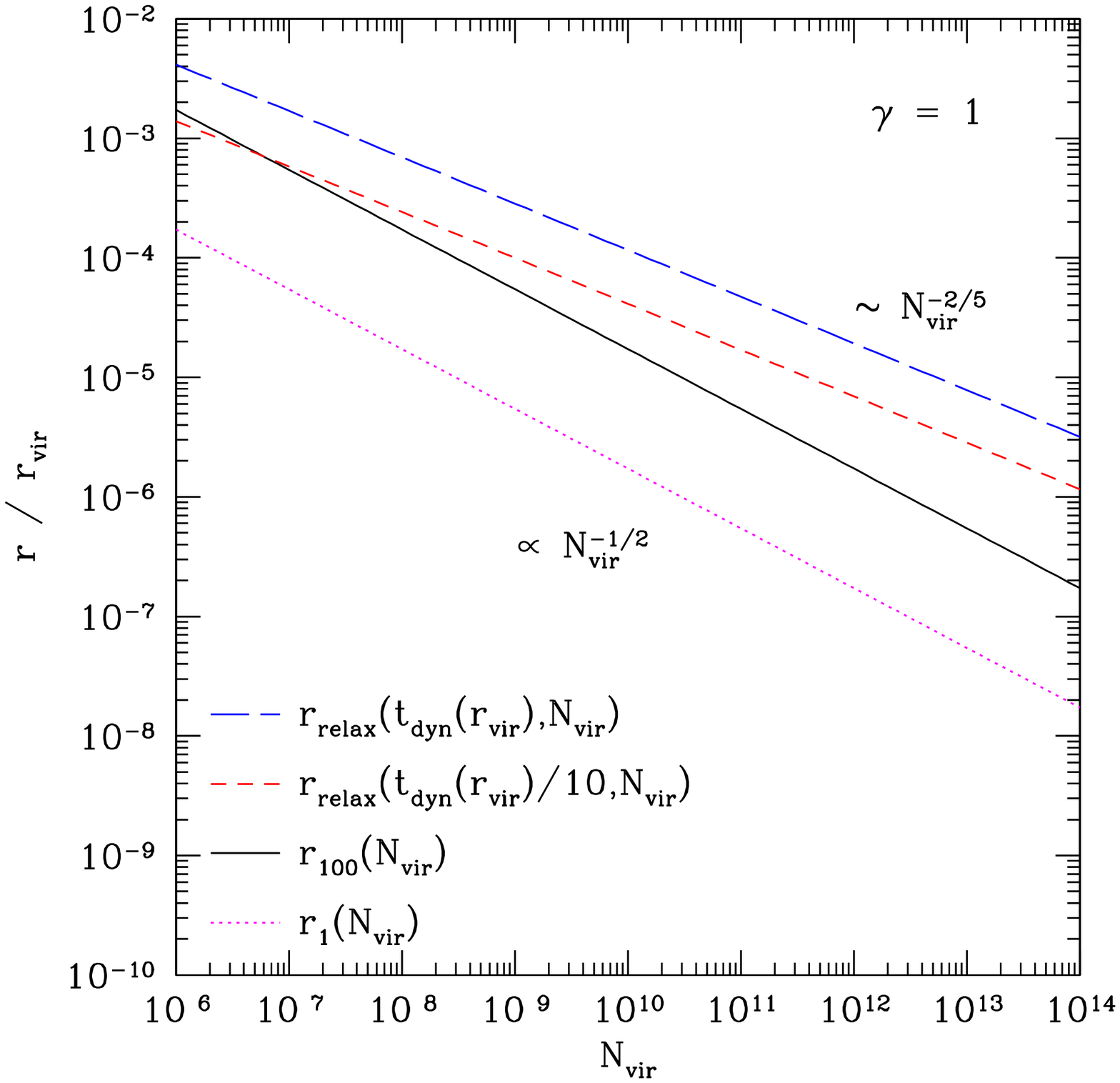}
	\includegraphics[width=0.49\textwidth]{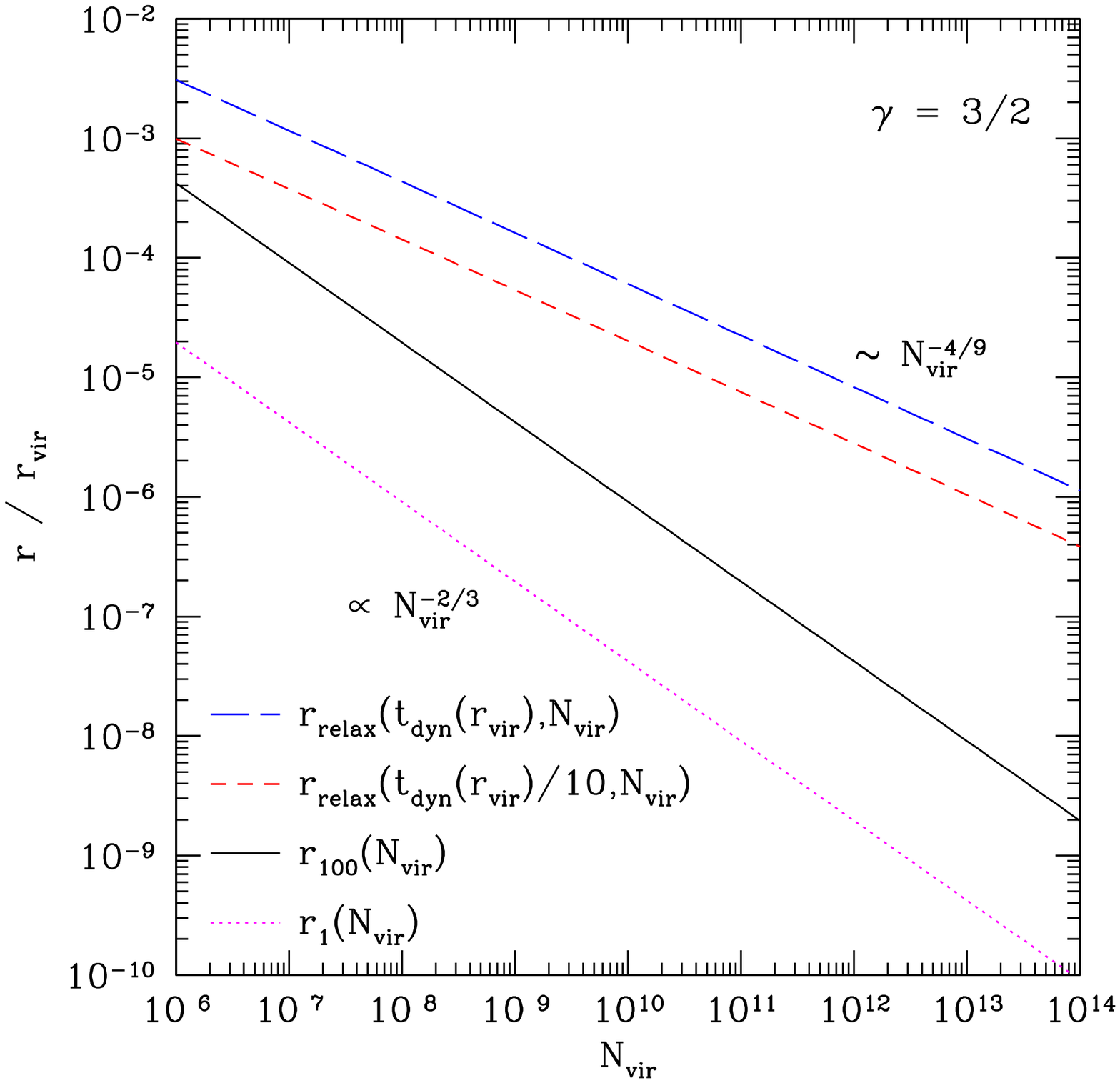}
	\caption{Plot of $r_1$, $r_{100}$, $\rrel(\tdyn(\rvir)/10)$, $\rrel(\tdyn(\rvir))$ as a function of $\Nvir$ for a spherical structure with $\cvir = 10$, $\alpha = 1$, $\beta = 3$ but different central profiles: $\gamma = 1$ (left) and $\gamma = 3/2$ (right). One can nicely see that in general the constraint on $\rres$ is set by the scale where relaxation becomes important since $\rres = \max(r_{100}(\Nvir),\rrel(t_0,\Nvir))$ for a given simulation time $t_0$. The asymptotic scaling for $r_N$ and $\rrel$ is given in the plots; the discrepancy between the sampled scale $r_{100}$ and the relaxation scale $\rrel$ as a function of $\Nvir$ is bigger for larger $\gamma$.}
	\label{fig:resolution}
\end{figure*}

We illustrate the resolution problem in more detail with a commonly used family of spherically symmetric density profiles: the so called $\alpha\beta\gamma$-models family \citep{1996MNRAS.278..488Z}. An $\alpha\beta\gamma$-model density profile is given by
\begin{equation} \label{eq:rho}
\rho(r) = \frac{\rho_0}{\left(\frac{r}{\rs}\right)^\gamma
\left(1+\left(\frac{r}{\rs}\right)^\alpha
\right)^{\left(\frac{\beta-\gamma}{\alpha}\right)}}~,
\end{equation}
where $\gamma$ determines the inner slope and $\beta$ the outer slope of the density profile, whereas $\alpha$ controls the transition between the inner and outer region. The normalisation is given by $\rho_0$ and $\rs$ is the scale radius defined by $\rs \equiv \rvir/\cvir$ ($\cvir$ is the virial concentration). Many models used by observers as well as theorists to describe structures in the universe belong to this family, e.g. the Hernquist profile \citep{1990ApJ...356..359H}, the NFW profile \citep{1996ApJ...462..563N} or the Moore profile \citep{1998ApJ...499L...5M} are just special cases of the general form (\ref{eq:rho}).

An obvious minimal criterion for a given length scale to be resolved is, that it has to be populated with enough particles by the sampling procedure. For example, if one would like to set up an isolated structure given by the density profile (\ref{eq:rho}), then the radius $r_{N}$ that contains $N$ particles is simply given by the solution of
\begin{equation} \label{eq:rNdef}
M(r_{N}) = N m~,
\end{equation}
where $M(r)$ is the enclosed mass given by
\begin{equation} \label{eq:Menc}
M(r) \equiv \int_0^r 4 \pi \rho(x) x^2 \d x~,
\end{equation}
and $m$ is the mass of the particles which can be expressed in virial quantities for single-mass models as $m = \Mvir / \Nvir$. In the central asymptotic regime (i.e. $r_{N} \ll \rs$), equation (\ref{eq:rNdef}) can be solved for $r_{N}$ to yield
\begin{equation} \label{eq:rN}
r_{N} = \left(\frac{3-\gamma}{4 \pi} \frac{N m}{\rho_0 \rs^{\gamma}}\right)^{\frac{1}{3-\gamma}}~.
\end{equation}
For $N = 1$ one obtains the scale of the location of the innermost particle, $\rimp = r_1$.\footnote{An alternative definition for the location of the innermost particle can be made with the mean particle separation given by $h(r) \equiv \sqrt[3]{\frac{m}{\rho(r)}}$. The two definitions are essentially equivalent and only differ by the geometric factor $\left(\frac{4 \pi}{3-\gamma}\right)^{\frac{1}{3-\gamma}}$ which is of order unity for most useful profiles.} But of course one particle is not enough to resolve that scale. If we say at least 100 particles are needed in the innermost bin so that the bin is well resolved, then $r_{100}$ provides a better estimate of the resolved scale.

Another constraint to resolution comes from the fact that often the structures that one would like to simulate can be very well approximated as collisionless systems, i.e. the local relaxation time-scale is much longer than the age of the system. This property must be preserved in $N$-body simulations of such collisionless systems. Otherwise, purely artificial relaxation processes due to under-resolving the structures will have a dominant effect on the dynamics of the system \citep{2003MNRAS.338...14P,2004MNRAS.348..977D,2004MNRAS.349.1117B}. This is not a numerical artefact since a real astrophysical system with a small particle number will also be subject to relaxation processes. The problem is, that in most cases, we can not simulate the astrophysical system under study with the number of particles the system would have in nature, e.g. a galaxy size dark matter halo would have of the order of $O(10^{67})$ dark matter particles if we assume that the dark matter particle has a mass of 100 GeV/c$^2$. Therefore, this effect due to under-resolving the system with not enough particles will always be a limitation of collisionless $N$-body simulations.

In principle, there are only few known dynamically stable systems in the universe, e.g. the Keplerian 2-body system or the figure-8 rotation of 3 bodies \citep{1993PhRvL..70.3675M, 2000AM....152..881C}.\footnote{This is only true in the Newtonian regime. In General Relativity, not even a 2-body orbit is dynamically stable and the orbit decays due to the emission of gravitational radiation \citep{1979RvMP...51..447D, 1977QJRAS..18....3I, 1997RvMP...69..337A}.} Dynamical effects like relaxation or evaporation will sooner or later lead to the disruption of any system. The issue is always on what time-scale the system is actually stable. The time-scale of interest is normally of the order of the age of the universe and for most cases collisionless astrophysical systems can be regarded as perfectly stable within that time-scale. This is different in $N$-body simulations. Here, one just tries to generate models that show the desired stability behaviour during the time-scale of interest with the least amount of particles needed. Artificial $N$-body models will be subject to such disruption effects much sooner than the real astrophysical system one wants to study.

We illustrate this in more details now. The local relaxation time is defined by
\begin{equation} \label{eq:trelax}
\trel(r) \equiv \frac{N(r)}{\ln(N(r))}~\tdyn(r)~,
\end{equation}
where
\begin{equation} \label{eq:tdyn}
\tdyn(r) \equiv 2 \pi \sqrt{\frac{r^3}{G M(r)}}
\end{equation}
is the dynamical or orbital time at radius $r$ and $N(r) \equiv M(r)/m$ denotes the number of particles within $r$. Here, we have a slightly different normalisation than in the usual expression for the local relaxation time since we dropped the factor of 8 in front of the logarithmic term in the denominator. We found better agreement with results from $N$-body simulations with this normalisation (see section \ref{chap:tests} below for more details). Had we kept the factor of 8, then our definition (\ref{eq:trelax}) for spherically symmetric structures would be equivalent to the empirical expression found by \citet{2003MNRAS.338...14P}. Normally, the Coulomb logarithm is given by $\ln(\Lambda) = \ln(b_{\mathrm{max}}/b_{\mathrm{min}})$, where $b_{\mathrm{max}}$ and $b_{\mathrm{min}}$ are the maximum and minimum impact parameters of the particles under consideration. Since the minimum impact parameter is related to the softening length and the latter scales with the number of particles, we prefer the direct formulation of the Coulomb logarithm as a function of the number of particles, $\ln(N(r))$.

In a simulation run for a time $t_0$ relaxation processes become important on a scale $\rrel$ given by the solution of
\begin{equation} \label{eq:rrelaxdef}
\trel(\rrel) = t_0~.
\end{equation}
In the central asymptotic regime (i.e. $\rrel \ll \rs$), equation (\ref{eq:rrelaxdef}) can be inverted for $\rrel$ to yield
\begin{equation} \label{eq:rrelax}
\rrel(t_0) =
\left(
\frac{W_{-1}(X)~\frac{(3 - \gamma) m}{4 \pi \rho_0 \rs^{\gamma}}}
{-\frac{1}{2} \frac{6-\gamma}{3-\gamma} \frac{\pi}{t_0} 
\left( \frac{3-\gamma}{G \pi \rho_0 \rs^{\gamma}} \right)^{\frac{1}{2}}} 
\right)^{\frac{2}{6-\gamma}}
\end{equation}
where $W_{-1}$ denotes the $k = -1$ branch of the Lambert W function \citep{1758AH......3..128L,1996ACM.....5..329C} and
\begin{equation}
X \equiv -\frac{1}{2} \frac{6-\gamma}{3-\gamma} \frac{\pi}{t_0} 
\left(\frac{3-\gamma}{G \pi \rho_0 \rs^{\gamma}}\right)^{\frac{1}{2}} 
\left( \frac{(3 - \gamma) m}{4 \pi \rho_0 \rs^{\gamma}} \right)^{\frac{\gamma}{2 (3-\gamma)}}~.
\end{equation}

In Fig. \ref{fig:resolution} we plot $r_1$, $r_{100}$, $\rrel(\tdyn(\rvir)/10)$ and $\rrel(\tdyn(\rvir))$ as a function of $\Nvir$ by setting $m = \Mvir / \Nvir$. The virial radius $\rvir$ is defined so that the enclosed average density within $\rvir$ is given by 
\begin{equation}
\frac{\Mvir}{4 \pi \rvir^3/3} = \rhovir = \Deltavir \rho_{\mathrm{crit},0} = 1.41\times10^{4}~\Mo~\kpc^{-3}
\end{equation}
where $\Deltavir = 178~\Omega_{\mathrm{M},0}^{0.45} \approx 104$ \citep{1998ApJ...503..569E} for our choice of cosmology with $\Omega_{\mathrm{M},0} = 0.3$, $\Omega_{\Lambda,0} = 0.7$, $\rho_{\mathrm{crit},0} = {3 H_0^2}/{8 \pi G}$ and $H_0 = 70~\km~\s^{-1}~\Mpc^{-1}~(h_0 = 0.7)$, which we use throughout this paper. The dynamical time at the virial radius is then
\begin{equation}
\tdyn(\rvir) = \sqrt{\frac{3 \pi}{G \Deltavir \rho_{\mathrm{crit},0}}} = 12.2~\Gyr~.
\end{equation}
For example for a galaxy size dark matter halo with $\Mvir = 10^{12}/h_0~\Mo = 1.43 \times 10^{12}~\Mo$ one would obtain $\rvir = 289~\kpc$ and the values for systems of different virial mass $\Mvir$ can be obtained by the simple scaling relation
\begin{equation}
\rvir = 289~\kpc \sqrt[3]{\frac{\Mvir}{1.43 \times 10^{12}~\Mo}}~.
\end{equation}

Although the virial radius is a rather artificial definition of the size of a dark matter halo and the virialised region of haloes in cosmological $N$-body simulations is generally much larger \citep{2006ApJ...645.1001P,2007ApJ...667..859D} it is a convenient normalisation and cut-off scale for isolated models since we are anyway mainly interested in the central dynamics of the structure. We set $\cvir = 10$, $\alpha = 1$, $\beta = 3$ in both cases and present plots for $\gamma = 1$ (left) respectively $\gamma = 3/2$ (right).  

We only plot these quantities for $\Nvir \geq 10^{6}$ since the expressions (\ref{eq:rN}) and (\ref{eq:rrelax}) are only valid in the central asymptotic regime. One can see that $r_N \propto \Nvir^{-\frac{1}{3-\gamma}}$ and $\rrel \sim \Nvir^{-\frac{2}{6-\gamma}}$ since the Lambert W function only varies very slowly as a function of $\Nvir$ which reflects the weak dependence of the logarithmic term of the local relaxation time. For a given particle resolution $\Nvir$ and simulation time $t_0$, the radius $\rres$ that we can still resolve with correct collisionless physics (i.e. this scale does not suffer from too much artificial relaxation) is given by 
\begin{equation}
\rres = \max(r_{100}(\Nvir),\rrel(t_0,\Nvir))~.
\end{equation}
As one can see from Fig. \ref{fig:resolution}, this resolution scale $\rres$ is in general set by $\rrel(t_0,\Nvir)$ for isolated high resolution structures. It is worth remarking here, that for structures assembled hierarchically in a cosmological $N$-body simulation, the amount of relaxation is significantly larger and $\rrel$ scales slower as a function of $\Nvir$ \citep{2004MNRAS.348..977D}. Although a structure might be sampled with enough particles at a certain scale at the final time, the relaxation time at that scale would have been much smaller in the past since particles were in lower mass structures during the hierarchical growth.

By inspecting Fig. \ref{fig:resolution}, we see that more than approximately of order $O(10^{12})$ particles in the centre of a structure with $\gamma = 1$ are needed in order to resolve scales of $\approx 10^{-5}~\rvir$. It is worth remarking, that with the same number of particles $\Nvir$ much smaller scales are populated in the $\gamma = 3/2$ profile than in the $\gamma = 1$ profile. Generally, the steeper the central profile, the more the particles are concentrated. But unfortunately, the relaxation scale $\rrel$ does not scale equally fast so that the discrepancy as a function of $\Nvir$ becomes bigger for larger values of $\gamma$.

Nonetheless, such an enormous amount of particles per structure is hardly doable today - even with large supercomputers. But since we only need this high resolution at the very centre of the structure, our solution to this problem is to use models where we only populate regions of the phase space that are in the centre or will reach the centre in the future with high resolution particles.

In section \ref{chap:method} we present the simple idea behind the multimass models and present stability tests in section \ref{chap:tests}. In section \ref{chap:preservation}, we test Dehnen's prediction with high resolution mergers and we summarize our results in section \ref{chap:conclusions}.

\section{Method} \label{chap:method}

\subsection{General model characteristics}

We restrict ourself to models of the form given by equation (\ref{eq:rho}) with $\gamma < 3$ for the mass not to diverge in the centre. Similarly, for $\beta \leq 3$ the total mass would diverge and we need to introduce a cut-off radius. We chose the form
\begin{equation} \label{eq:rhocut}
\rho(r) = \left\{
\begin{array}{ll}
\frac{\rho_0}{\left(\frac{r}{\rs}\right)^\gamma \left(1+\left(\frac{r}{\rs}\right)^\alpha
\right)^{\left(\frac{\beta-\gamma}{\alpha}\right)}}	&
r \leq \rcut \\
\rho(\rcut) \left(\frac{r}{\rcut}\right)^\delta \mathrm{e}^{-\frac{r-\rcut}{\rdec}} &
r > \rcut
\end{array}
\right.
\end{equation}
where $\delta$ and $\rdec$ are free parameters \citep{1999MNRAS.307..162S, 2006ApJ...641..647K}. By requiring the logarithmic slope to be continuous at $\rcut$, we get
\begin{equation}
\delta = \frac{\rcut}{\rdec} - \frac{\gamma + 
\beta \left(\frac{\rcut}{\rs}\right)^{\alpha}}{1 + \left(\frac{\rcut}{\rs}\right)^{\alpha}}~.
\end{equation}
We set the truncation scale $\rdec = 0.3~\rcut$ in order not to make the truncation too sharp. A too sharp truncation can lead to an instability of the model around $\rcut$ as seen in \cite{2003DiplT.........Z}. For $\beta > 3$ we simply set $\rcut = \infty$ (i.e. no cut-off) while for $\beta \leq 3$ one has to specify a cut-off scale $\rcut$ (e.g. $\rcut = \rvir$). By further specifying $\rs$ and $M(\rcut)$, the normalisation $\rho_{0}$ is given by
\begin{equation}
\rho_0 = \frac{M(\rcut)}{4 \pi \rs^3 I_{\mathrm{M}}}
\end{equation}
where
\begin{equation}
I_{\mathrm{M}} \equiv \int_0^{q} \frac{x^{2-\gamma}}{\left(1+x^\alpha\right)^{\left(\frac{\beta-\gamma}{\alpha}\right)}} \d x
\stackrel{q=\infty}{=} \frac{\Gamma\left(\frac{\beta-3}{\alpha}\right) \Gamma\left(\frac{3-\gamma}{\alpha}\right)}{\alpha \Gamma\left(\frac{\beta-\gamma}{\alpha}\right)}
\end{equation}
with $q = \rcut / \rs$ and $\Gamma$ is the standard gamma function.

\subsection{Distribution function as probability density function}

For spherical systems with an isotropic velocity distribution one can calculate the distribution function, which in that case only depends on energy, by the Eddington inversion \citep{1987gady.book.....B}. Hence, by restricting to models with an isotropic velocity distribution, we can calculate the distribution function for our spherical structure models described by equation (\ref{eq:rhocut}) - at least numerically. Since the state of a system at a given time is completely described by the distribution function $f(\vec{r},\vec{v})$, we use it as a probability density function in order to sample the phase space with particles,
\begin{equation}
p_{\mathrm{6D}}(\vec{r},\vec{v}) \d \vec{r} \d \vec{v} = \frac{f(\vec{r},\vec{v})}{\Mtot} \d \vec{r} \d \vec{v}
\end{equation}  
is the probability that a particle is in the volume $\d \vec{r} \d \vec{v}$ around the phase space point $(\vec{r},\vec{v})$. By integrating out the velocities and using spherical symmetry, we get for the probability density $p(r)$ in coordinate space
\begin{equation}
p(r) \d r = \frac{4 \pi r^2 \rho(r)}{\Mtot} \d r
\end{equation}
and the positions can now be sampled by using the quantile function, which is the inverse of the cumulative probability distribution function $M(r)/\Mtot$ for the above probability density function $p(r)$. For a particle at location $\vec{r}_i$ we get now the following probability density for the magnitude of the velocities 
\begin{equation}
p(r_i,v) \d v = \frac{4 \pi v^2 f(r_i,v)}{\rho(r_i)} \d v
\end{equation}
where $r_i = |\vec{r}_i|$ and isotropy in velocity space was used. In general, the distribution function can only be calculated numerically for the large family of models described by the density profile (\ref{eq:rhocut}) although a few analytical solutions are known \citep[e.g.][]{1990ApJ...356..359H,1993MNRAS.265..250D,1994AJ....107..634T}. Hence, numerical integration and inversion for $p(r_i,v)$ in order to calculate the quantile function is difficult and one generally uses the acceptance-rejection technique \citep{1951NBSAM..12...36v,1994MNRAS.269...13K} for the Monte Carlo sampling of the velocities.

This Monte Carlo sampling procedure directly from the distribution function $f(\vec{r},\vec{v})$ leads to perfectly stable equilibrium models as was shown in \cite{2004ApJ...601...37K} for single-mass models. These models do not show the flattening in the central part of the structure during evolution as it is obtained in the case of the assumption of a local Maxwellian velocity distribution with the velocity dispersion given by the Jeans equation.

\subsection{Multimass refinement}

Since in general we only need the high resolution sampling for structures in the central region, we use a multimass technique to accomplish that. The general idea is that the central region is populated by lighter particles and the outer parts of the structure are sampled by heavier particles. In addition, a refinement depending on the orbit of the particle is applied in order to minimise the perturbations of heavy particles from the outer parts of the structure to the high resolution centre.

\subsubsection{Shell refinement}

By specifying an inner shell radius $\rsi$ and a number of particles $N_0$ within that radius, the mass of the particles in the central sphere is simply $m_0 = M(\rsi) / N_0$ and sets the effective central resolution of the structure. One can further choose an outer shell radius $\rso$, the number of shells $\Nshell$ between $\rsi$ and $\rso$, and the mass ratio between the particle mass in neighbouring shells $\Delta \RM$. Since one has to split particles if one uses the orbit dependent refinement (see below for more details) and we do not want to have particles with masses smaller than $m_0$, $\Delta \RM$ has to be a natural number, i.e. $\Delta \RM \in \mathbb{N}$. The resulting structure (without orbit dependent refinement) has then a total of $\Nshell + 2$ different mass species with $m_i = m_0 (\Delta \RM)^i, i = 0\ldots\Nshell+1$. The shells are equally spaced in logarithmic intervals which determines the number of particles within each shell.

Of course this technique introduces some further numerical artefacts as e.g. heating of the light particles by the heavy particles or mass segregation of the heavy outer particles during time evolution. The dynamical friction force that a particle of mass $M$ experiences in a homogeneous sea of light particles with mass $m \ll M$ is $F_{\mathrm{df}} \propto M^2$ and the time-scale for this particle to reach the centre of the structure is $t_\mathrm{df} \propto M^{-1}$ \citep{1987gady.book.....B}. Hence, by choosing moderate mass ratios $\Delta \RM$ between neighbouring shells, we can reduce the effect of mass segregation. By scaling the softening lengths of the heavy particles as a function of their mass, artificial 2-body scattering can be reduced. The actual role of the softening length is a cut-off scale for the singularities introduced by the Monte Carlo sampling \citep{1993MNRAS.262.1013L}. A natural inner scale for a given density profile and mass resolution is set by the location of the innermost particle $\rimp = r_1 \propto m^{\frac{1}{3-\gamma}}$ (see equation (\ref{eq:rN})). Hence, we chose to scale the softening like $r_1(m_i)$ for each species, resulting in
\begin{equation} \label{eq:softscale}
\epsilon_i = \epsilon_0 \left(\frac{m_i}{m_0}\right)^{\frac{1}{3-\gamma}} = \epsilon_0 (\Delta \RM)^{\frac{i}{3-\gamma}}~.
\end{equation}
This can be seen as a generalised scaling for arbitrary central profiles $\gamma$ in isolated structures of the often used rule for cosmological $N$-body simulations, where one scales the softening length by $\epsilon_i \propto \sqrt[3]{m_i}$. Of course, for $\gamma = 0$, which would correspond to the homogeneous case, the two scaling relations are identical.

\subsubsection{Orbit dependent refinement}\label{chap:odr}

In addition, we refine particles in the outer parts of the structure depending on their orbit. From the initial position and velocity, we can calculate the pericentre distance $r_{\mathrm{peri},k}$ of a specific particle in the smooth potential given by the density profile. By choosing the maximum orbital refinement radius $\rmor$ we determine the split factor $f_k$ for that particle by
\begin{displaymath}
f_k = \left\{
\begin{array}{ll}
\frac{m_k}{m_0}	& r_{\mathrm{peri},k} \leq \rsi \\
1 & \rmor \leq r_{\mathrm{peri},k} \\
\frac{m_k}{m_0}	+ \frac{\log\left(r_{\mathrm{peri},k}/\rsi\right)}{\log\left(\rmor/\rsi\right)}\left(1-\frac{m_k}{m_0}\right) &
\rsi < r_{\mathrm{peri},k} < \rmor < r_i \\
\frac{m_k}{m_0}	+ \frac{\log\left(r_{\mathrm{peri},k}/\rsi\right)}{\log\left(r_i/\rsi\right)}
\left(1-\frac{m_k}{m_0}\right) &
\rsi < r_{\mathrm{peri},k} \leq r_i \leq \rmor \\
\end{array}
\right.
\end{displaymath}
\begin{equation}
\end{equation}
where $r_i$ is the outer boundary of shell $i$ that contains the particle with index $k$ and is given by
\begin{equation}\label{eq:ri}
r_i = \rsi \left(\frac{\rso}{\rsi}\right)^{\frac{i}{\Nshell}}
\end{equation}
for $0 \leq i \leq \Nshell$ and $r_{\Nshell+1} \equiv \infty$. If this factor $f_k > 1$, then we replace that particle with $f_{\mathrm{split},k} = \mathrm{int}(f_k)$ copies of mass $m_{\mathrm{split},k} = m_k / f_{\mathrm{split},k}$ randomly placed on a sphere of radius $d_k$, which is the distance of the original particle to the geometric centre. Here, $\mathrm{int}(x)$ is the function that rounds the real value $x$ to its nearest integer. We split the velocity of the original particle into a radial and tangential part. The new particles will have the same radial velocity component as the original particle but a new, random tangential component of the same magnitude as the original one. With such a splitting procedure, we keep the velocity configuration of the structure, i.e. the total kinetic energy is the same and the new particles are on the same orbit as the original particle. Of course, we also scale the softening of the new particles according to equation (\ref{eq:softscale}). With this orbit depending refinement technique we minimise the perturbation of the high resolution central region by heavy particles from the outer region, e.g. a heavy particle on a perfectly radial orbit would be split up into copies that have the same mass as the particles in the central high resolution region. Therefore, this procedure generates a protection layer around the high resolution region and very few heavy particles will diffuse into that region.

\subsubsection{Choice of parameters - some guidelines}

We can express the mass of a particle in shell with index $i$ before the orbit dependent refinement as a function of radius
\begin{equation}
m_i = m_0 \left(\frac{r_i}{\rsi}\right)^{\kappa}
\end{equation}
where $r_i$ is the outer boundary of shell $i$ and
\begin{equation}
\kappa = \Nshell \frac{\log\left(\Delta \RM \right)}{\log\left(\rso/\rsi\right)}~.
\end{equation}
The mass in each shell $M_i$ scales like the enclosed mass in the central asymptotic regime, i.e. $M_i \propto r^{3-\gamma}$. Therefore, a choice of $\kappa = 3-\gamma$ results in an equal number of particles in each shell before orbit dependent refinement. Equation (\ref{eq:rrelax}) suggests, that one should scale $m \sim r^{\frac{6-\gamma}{2}}$ in order to keep the local relaxation time constant, which would be the most efficient distribution of particles. But a steeper scaling with $\kappa > 3-\gamma$ also means that shells in the outer part might have just a few very heavy particles which does not give a good sampling of the structure. This behaviour becomes worse since the asymptotic scaling of $M_i \propto r^{3-\gamma}$ is only valid in the asymptotic central regime and becomes flatter in the outer parts of the density profile. Additionally, most of the work load is soon dominated by the high resolution centre, so that even fewer particles in the outer part does not result in a significant computer run time gain (see e.g. Fig. \ref{fig:speed}).

The value of $\kappa$ is degenerate in the sense that many different choices of $\Nshell$, $\rsi$, $\rso$ and $\Delta \RM$ can give the same value for $\kappa$. Here we give some guidelines on how to chose the different parameters. Unfortunately, there is not a set of universal parameters that works for all possible models within the $\alpha\beta\gamma$-family. It often also depends on the specific needs of the simulation. In general, we prefer not too big mass ratios between neighbouring shells in order to keep mixing and mass segregation effects to a minimum. This means that in most cases we set $\Delta \RM = 2$. The inner shell radius $\rsi$ is set so that the central region of interest is well sampled with high resolution particles. The outer shell radius $\rso$ should not be chosen too large, i.e. on scales where one is far from the central asymptotic scaling of the enclosed mass. We made good experience with choosing $\rso \le \rvir$. With the final degree of freedom, the $\Nshell$ parameter, we control $\kappa$ so that $\kappa$ is around $3-\gamma$. In order to protect the central high resolution region from heavy particles, one has to chose $\rmor$ large enough. We recommend a value of $\rmor \ge 10~\rsi$. The chosen values of the stable test models in section \ref{chap:tests} can also help to guide the reader choosing the parameters.

Hence, with a careful choice of parameters, perturbations of structures can be minimised and effects on global characteristics like the radial density profile are small. The multimass technique is therefore an efficient method to perform high resolution $N$-body simulations.

\section{Tests} \label{chap:tests}

\subsection{Two-shell models without orbit dependent refinement}

\begin{figure*}
	\centering
		\includegraphics[width=0.495\textwidth]{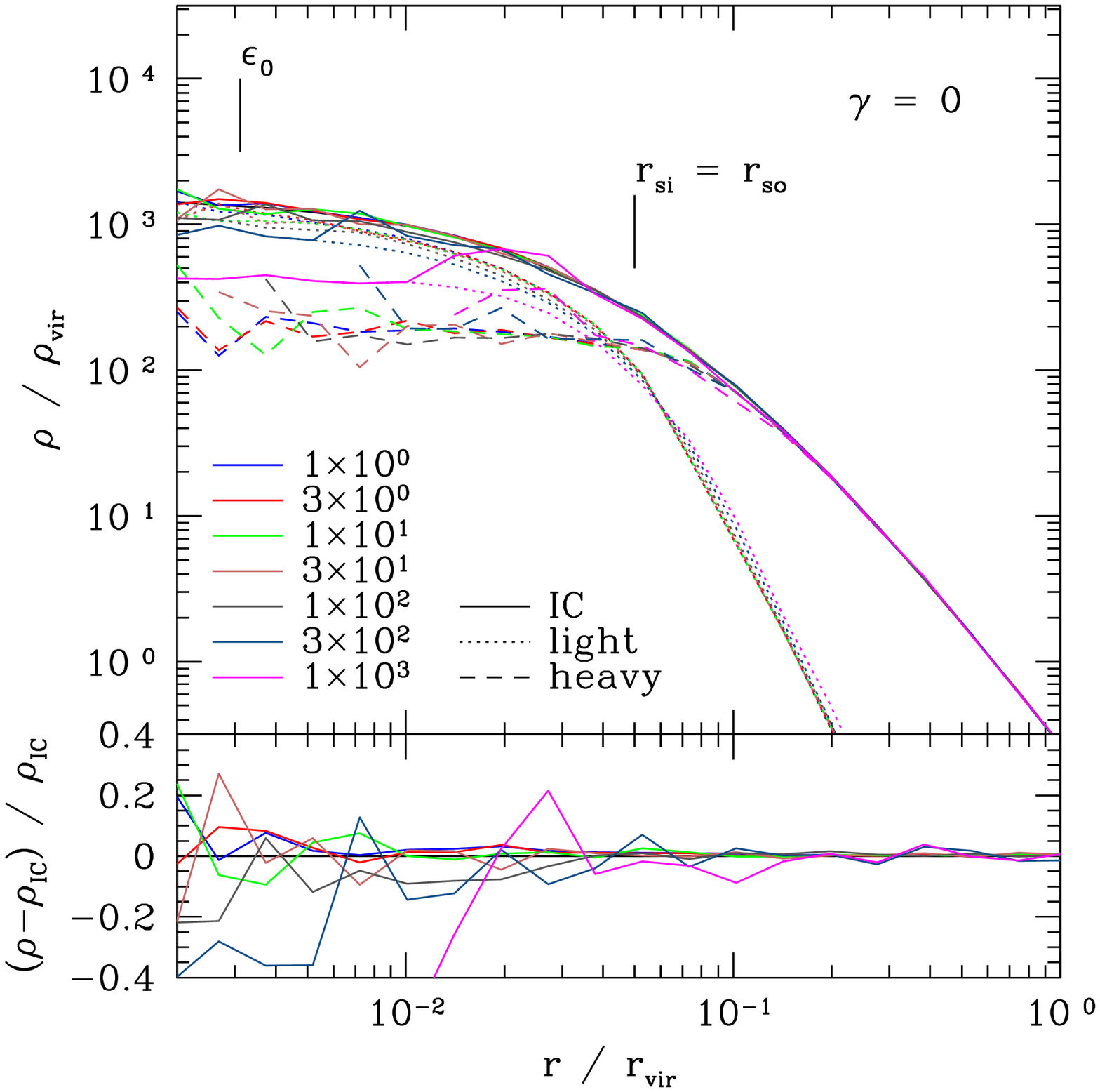} \hfill
		\includegraphics[width=0.495\textwidth]{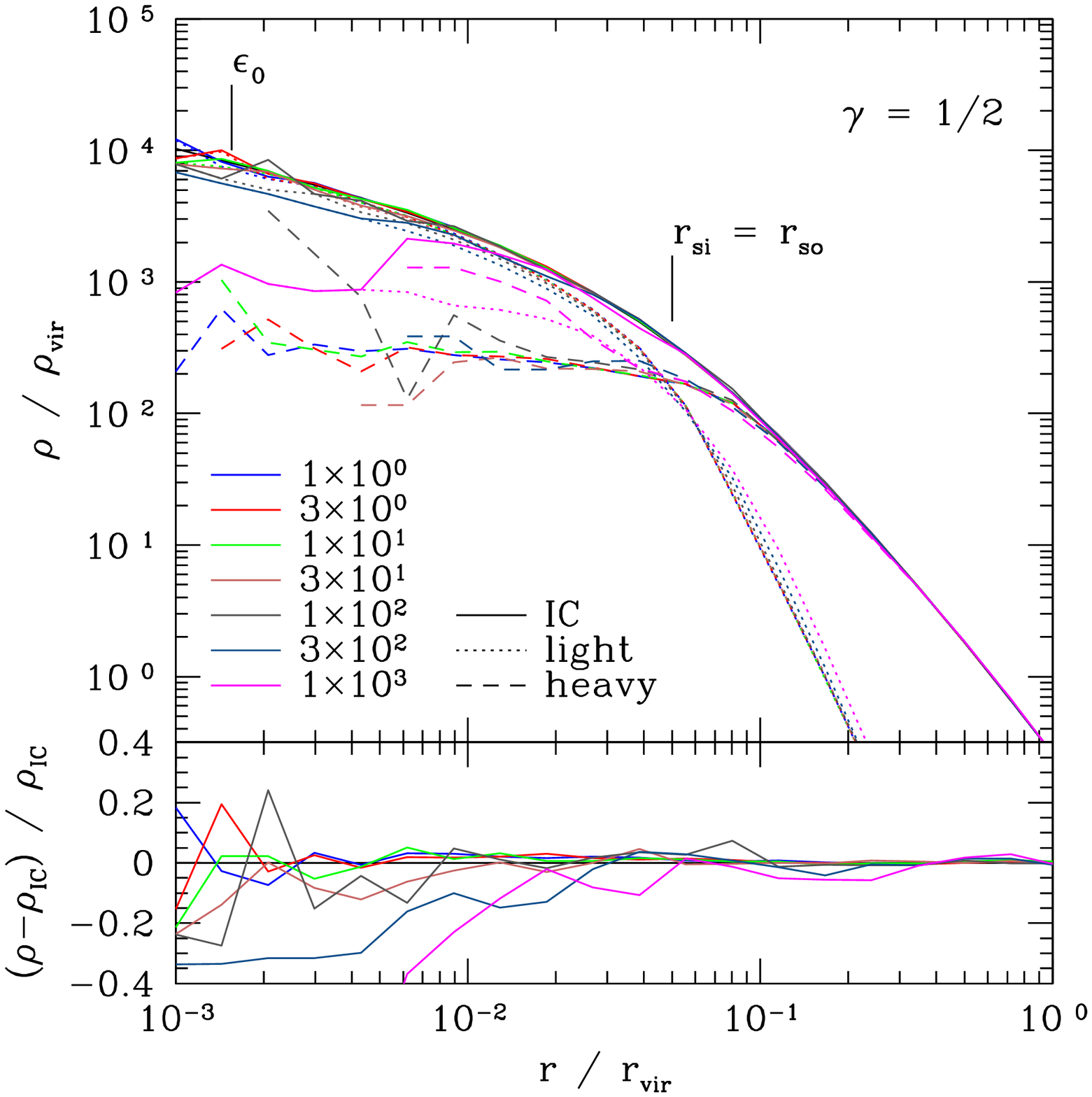} \\
		\includegraphics[width=0.495\textwidth]{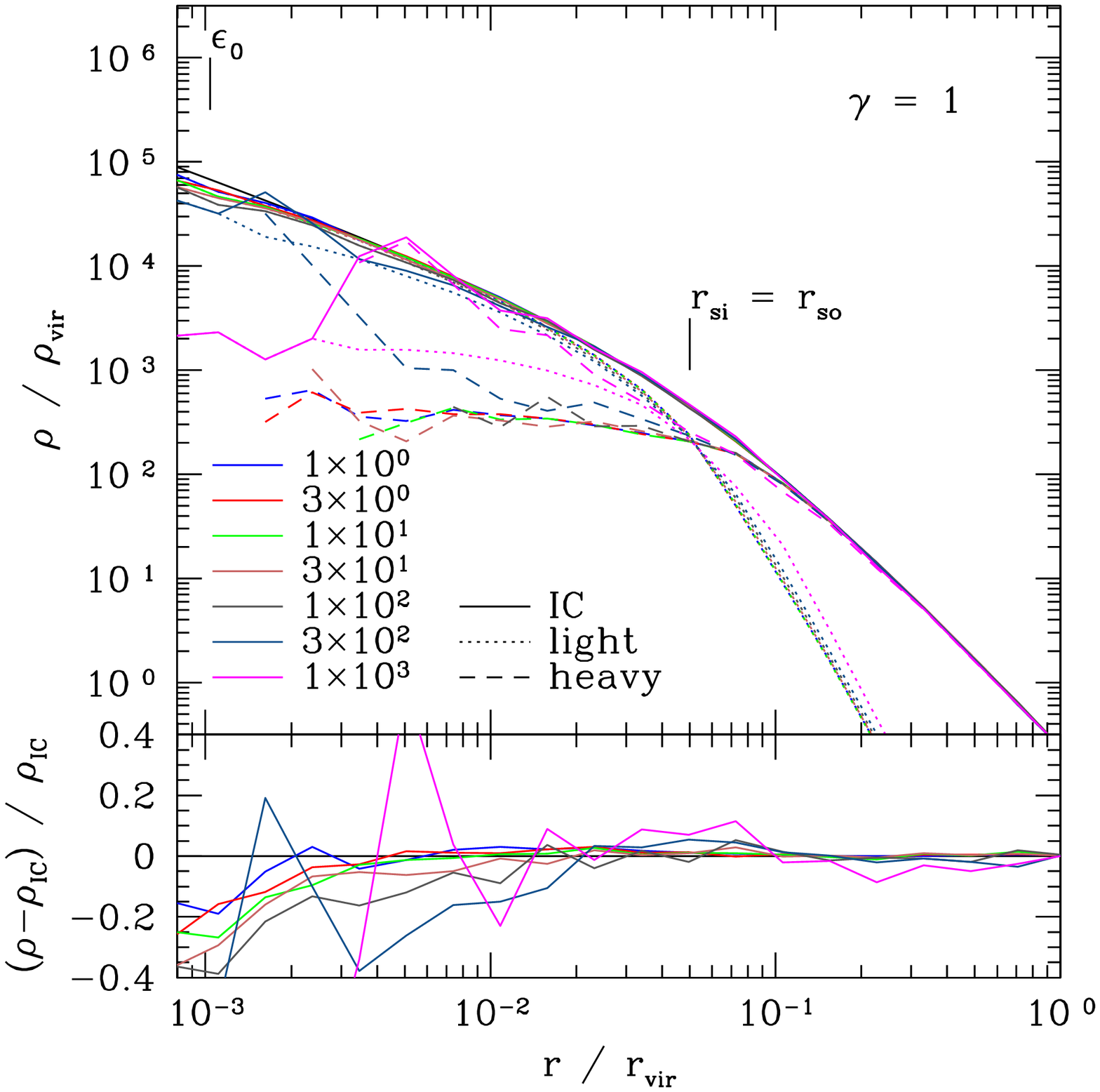} \hfill
		\includegraphics[width=0.495\textwidth]{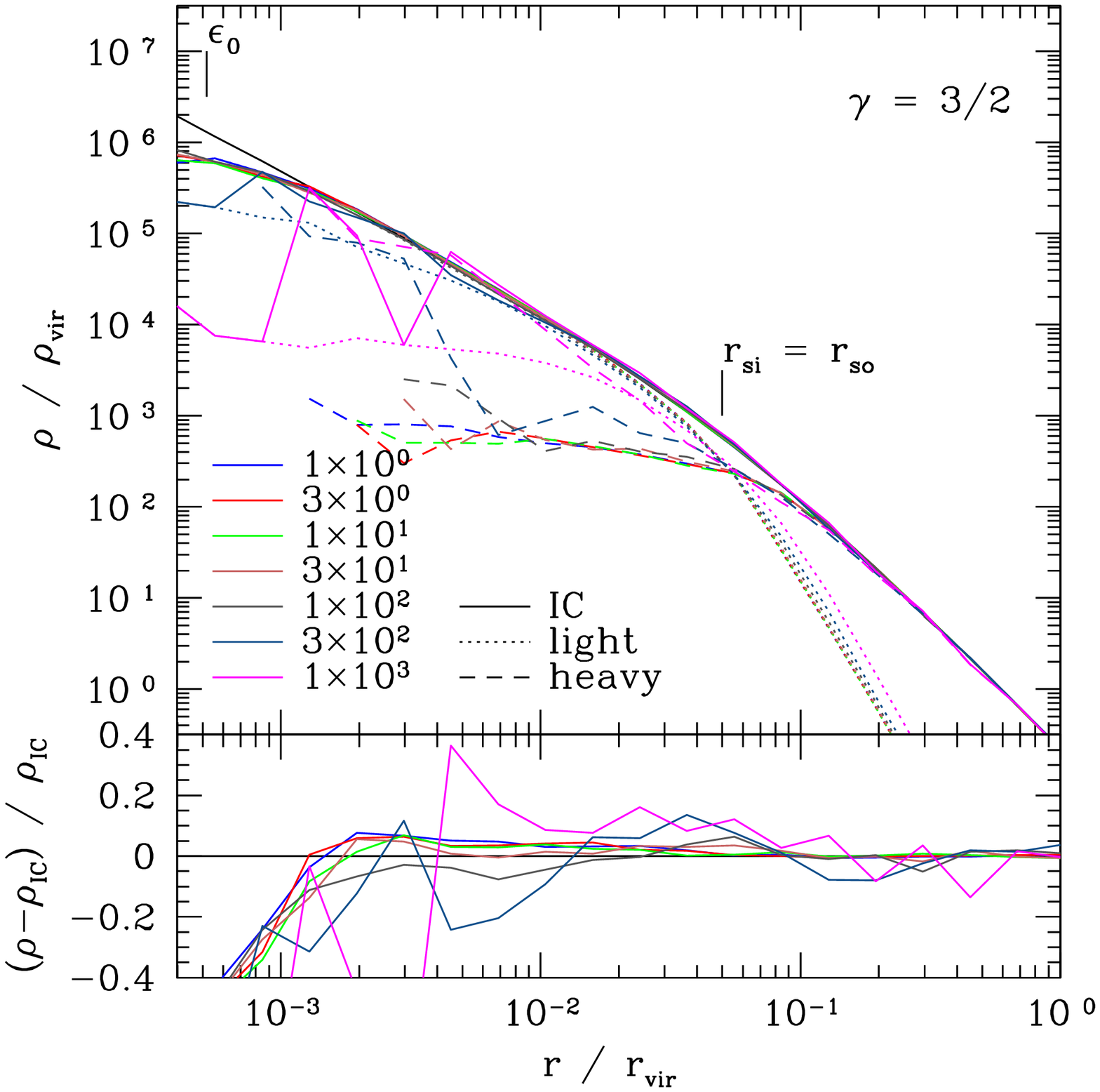} \\
		\caption{Multimass stability tests for mass profiles with different inner slope $\gamma = 0 \ldots 3/2$. Each run was evolved for 10 Gyr. The top panels for each slope $\gamma$ show the total density profile and the sub-profiles [(light,heavy) = (dotted,dashed)] for the different mass ratio runs normalised to the virial density $\rhovir = 1.41 \times 10^{4}~\Mo~\kpc^{-3}$. The lower panel shows the relative change of the total profile with radius $(\rho-\rho_{\mathrm{IC}})/\rho_{\mathrm{IC}}$ normalised to the profile of the initial conditions $\rho_{\mathrm{IC}}$. The total density profiles are stable for moderate mass ratios $\Delta \RM$ and only show very small perturbations. For too high mass ratios, the total density profile is stronger perturbed by heating effects of the heavy particles and the heavy particles sink to the centre due to the higher efficiency of dynamical friction.}
		\label{fig:shelltest}
\end{figure*}

\begin{table*}
	\caption{Summary of parameters for the different two-shell models.}
	\label{tab:models}
	\begin{tabular}{ccccc}
		\hline
		$\gamma$ & 0 & 1/2 & 1 & 3/2 \\
		\hline
		$\epsilon_0~[\rvir]$ & 
		$3.11 \times 10^{-3}$ & 
		$1.56 \times 10^{-3}$ & 
		$1.04 \times 10^{-3}$ & 
		$5.18 \times 10^{-4}$ \\
		$\Nvir$ &
		$7.21 \times 10^{6}$ &
		$4.88 \times 10^{6}$ &
		$3.25 \times 10^{6}$ &
		$2.11 \times 10^{6}$ \\
		$\rrel(10~\Gyr,\Nvir)~[\rvir]$ &
		$2.44 \times 10^{-3}$ & 
		$2.25 \times 10^{-3}$ & 
		$2.02 \times 10^{-3}$ & 
		$1.78 \times 10^{-3}$ \\		
		\hline
	\end{tabular}
	\medskip
\end{table*}

In order to illustrate the basic idea and the limitations of the multimass technique, we present a series of simple two-shell models without an orbit dependent refinement, i.e. $\rmor = 0$. In this series we chose four different profiles from the $\alpha\beta\gamma$-family described by equation (\ref{eq:rhocut}). The following density profile parameters were the same for all models: outer profile $\beta = 3$, transition coefficient $\alpha = 1$ and concentration $\cvir = 20$. We varied the inner profile form $\gamma = 0 \ldots 3/2$. For the refinement we chose $\rsi = \rso = \rs$, $\Nshell = 0$ (i.e. only two shells) and $N_0 = 3 \times 10^5$ for all models. For the different runs, we varied the mass ratio $\Delta \RM$ with the following values $1\times10^0, 3\times10^0, 1\times10^1, 3\times10^1, 1\times10^2, 3\times10^2, 1\times10^3$. The choice of softening for the high resolution particles $\epsilon_0$ and the total number of particles $\Nvir$ in the case of equal mass particles in the inner and outer shell, which corresponds to the effective resolution of the multimass models, as well as the estimated relaxation scale after 10 Gyr, $\rrel(10~\Gyr,\Nvir)$, are given in Table \ref{tab:models}. The softening lengths of the heavy particles were scaled as described by equation (\ref{eq:softscale}).

Each of these 28 models was evolved in isolation for 10 Gyr in order to test the stability of the structures. For the time evolution of all models we present in this paper, we use \textsc{pkdgrav}, a state-of-the-art tree code written by by Joachim Stadel \citep{2001PhDT........21S}. We used the dynamical time-stepping scheme developed by \citet{2007MNRAS.376..273Z}. For multimass models it is important to use a time-step for the particles that is based on the true dynamical time of a particle and independent of the softening because heavy particles with a large softening and light particles with a small softening can now mix. If for example two such particles were at the same geometric distance from the centre they should both take the same time-step. This is guaranteed by a criterion that is based on the dynamical time. Time-step criteria that are based on or scale with characteristics of the particle can't fulfil that and can lead to physically wrong time-steps. In addition, for high resolution models, the dynamical time-stepping is more efficient and more accurate in the high resolution centre. For a detailed discussion about time-stepping criteria see \citet{2007MNRAS.376..273Z}.

In Fig. \ref{fig:shelltest}, we present the density profiles of these runs after 10 Gyr. For moderate mass ratios $\Delta \RM$ up to 10-30 (or even $\Delta \RM \approx 100$ for steep central profiles) the effects on the total density profile are small and the profile remains stable down to the level of a few $\epsilon_0$. Such deviations are anyway expected since the forces in \textsc{pkdgrav} are softened if two particles have distances of order of their softening length. By comparing the equal mass cases for the two steepest central profiles, we see that the flattening effect due to relaxation sets in at a radius that is a little bit smaller than the estimated value $\rrel(10~\Gyr,\Nvir)$. Hence, our estimate (\ref{eq:rrelax}) is a good estimate. If we would keep the factor 8 in front of the logarithmic term for the relaxation scale, $\rrel$ would be approximately twice as large, confirming that it is rather a conservative estimate. This was our motivation to drop the factor of 8.

Figure \ref{fig:shelltest} also illustrates that the different species form stable sub-profiles. The shell radius was chosen in a zone where the local density profile is steep so that the transition region is small and in the inner or outer regions the light or heavy particles dominate respectively. Only for very high mass ratios, the total density profile is strongly perturbed by heating effects of the heavy particles and the heavy particles sink to the centre due to the higher efficiency of dynamical friction. But for moderate mass ratios, these effects are small and expected only for much longer time-scales.

\begin{figure}
	\centering
		\includegraphics[width=\columnwidth]{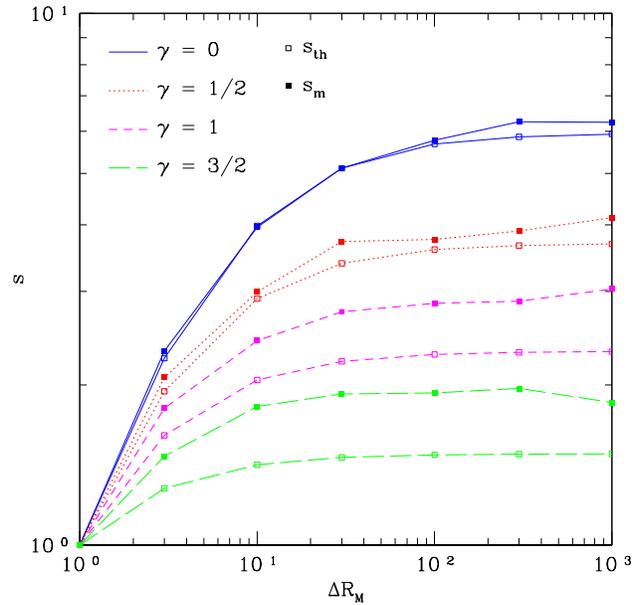}
		\caption{The theoretical and measured speed up factors $s_{\mathrm{th}}$ respectivley $s_{\mathrm{m}}$ as a function of mass ratio $\Delta \RM$ for the different central profiles $\gamma$ for the two-shell models without orbit dependent refinement. A substantial fraction of time is gained in all cases. Most of the time gain is already obtained for small mass ratios.}
		\label{fig:speed}
\end{figure}

The main advantage of these multimass models is the speed up gain. We can estimate a theoretical speed up factor $s_{\mathrm{th}}$ by the ratio of the number of force evaluations between a single-mass and a multimass model of the same density structure
\begin{eqnarray}
s_{\mathrm{th}} &\equiv& \frac{\int_{\rimp}^{\infty} \frac{4 \pi \rho(r) r^2}{m_0 \Delta T(r)} \d r}
{\sum_{i=0}^{\Nshell+1} \int_{r_{i-1}}^{r_i} \frac{4 \pi \rho(r) r^2}{m_i \Delta T(r)} \d r} \label{eq:sthc} \\
 & = & \frac{\sum_{j = 1}^{\Ntot} \frac{m_j}{m_0}\frac{1}{\Delta T(r_j)}}{\sum_{j = 1}^{\Ntot} \frac{1}{\Delta T(r_j)}} \label{eq:sthd}
\end{eqnarray}
where $r_i$ is the outer boundary of shell $i$ (see also equation \ref{eq:ri}) and we set $r_{-1} \equiv \rimp$ and $r_{\Nshell+1} \equiv \infty$. $\Delta T(r)$ is the time-step a particle takes at radius $r$. Here we use a dynamical time-stepping scheme where
\begin{equation}
\Delta T(r) \equiv \eta_{\mathrm{D}} \sqrt{\frac{r^3}{G M(r)}} = \frac{\eta_{\mathrm{D}}}{2 \pi} \tdyn(r)
\end{equation}
with $\eta_{\mathrm{D}} = 0.03$. For more details about the dynamical time-stepping scheme please consult \citet{2007MNRAS.376..273Z}. The second expression for $s_{\mathrm{th}}$ is valid for a discrete sampling of the structure and can be obtained from the first expression by plugging in the discrete density
\begin{equation}
\rho(r) = \sum_{j = 1}^{\Ntot} m_j \frac{\delta(r_j - r)}{4 \pi r^2}
\end{equation}
where we used the Dirac delta distribution in spherical coordinates.
 
In Fig. \ref{fig:speed}, we plot the theoretical estimate $s_{\mathrm{th}}$ and the measured speed up factor $s_{\mathrm{m}} \equiv T_0 / T$, where $T_0$ is the time needed by the equal mass run for the 10 Gyr simulation time stability test run described above and $T$ is the computer run time needed by the multimass model. Figure \ref{fig:speed} illustrates that we gain a substantial fraction of computer run time in all cases. Most of the gain is already obtained for small mass ratios $\RM$. For example, the run with inner slope $\gamma = 0$ and mass ratio $\Delta \RM = 10$ is approximately four times faster than the same run without multimass refinement and does not show any perturbation effect of the multimass technique on the density profile.

The steeper the central profile, the less is the computer run time gain. This is due to the fact that most of the work in the $N$-body simulation for steep profiles is concentrated in the centre. In the centre, the particles are on very small time-steps compared to the less dense, outer regions of a structure in an $N$-body simulation and as a consequence a lot of expensive force calculations are needed.

For flat density profiles the theoretical speed up estimate $s_{\mathrm{th}}$ agrees quite well with the measured value $s_{\mathrm{m}}$. Only for steeper central profiles the measured speed up factor $s_{\mathrm{m}}$ is larger than the theoretical prediction. This can be explained by the fact that in the steep profiles the particles in the very centre are expelled due to numerical effects (see for example the central density profiles in figure \ref{fig:shelltest}). As mentioned above, in steep central profiles the work is dominated by the central particles. Hence, if one looses some of them by numerical effects the work load decreases and the speed up factor becomes larger than the theoretical estimate which assumes that the density profile stays perfectly stable over time. In flatter central profiles this numerical effect is much weaker and the theoretical estimate agrees quite well with the measured speed up factor. In that sense, the theoretical estimate $s_{\mathrm{th}}$ is a lower limit for the speed up gain.

\subsection{Multimass models with orbit dependent refinement}

\begin{figure*}
	\centering
		\includegraphics[width=0.495\textwidth]{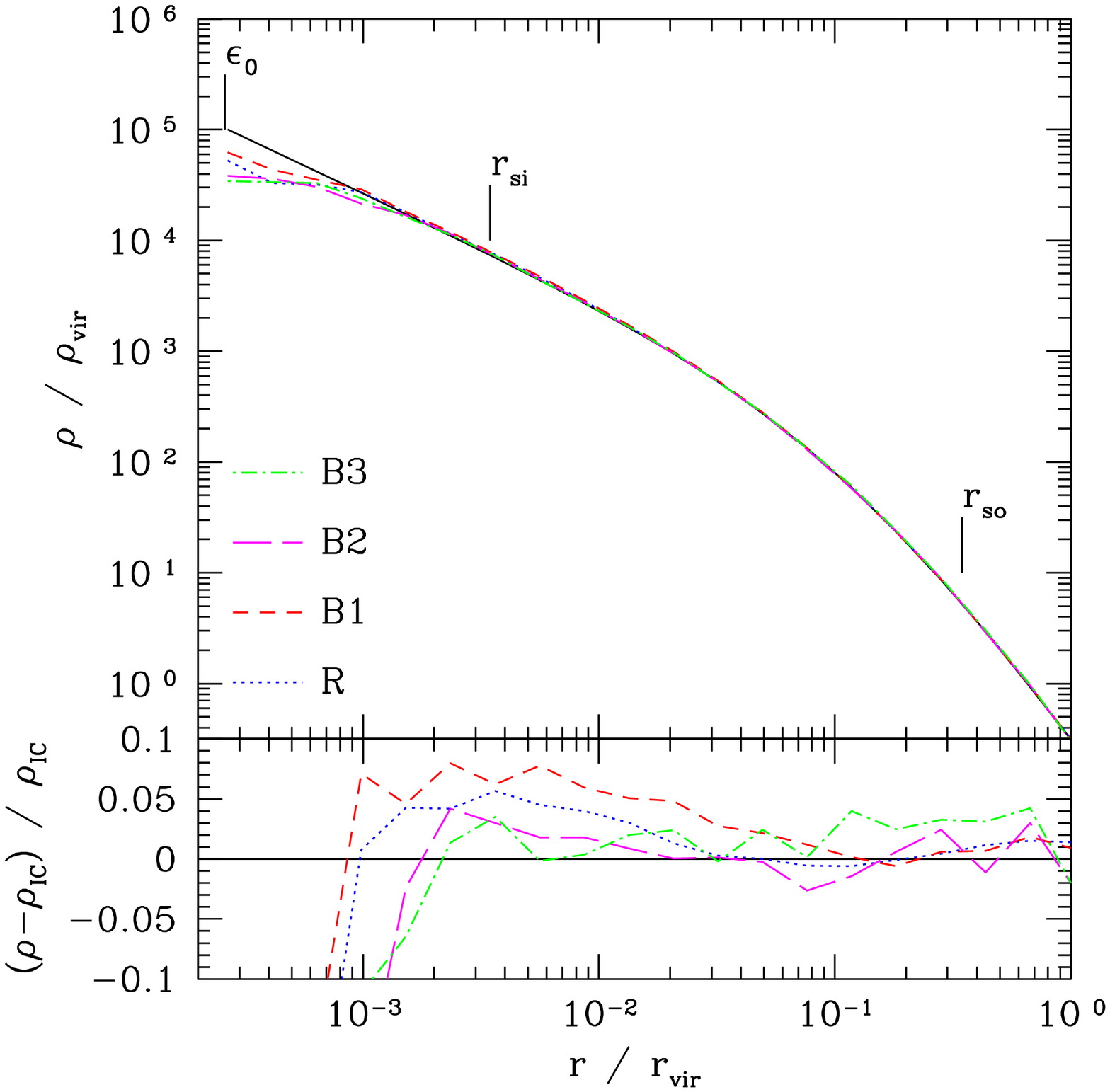} \hfill
		\includegraphics[width=0.495\textwidth]{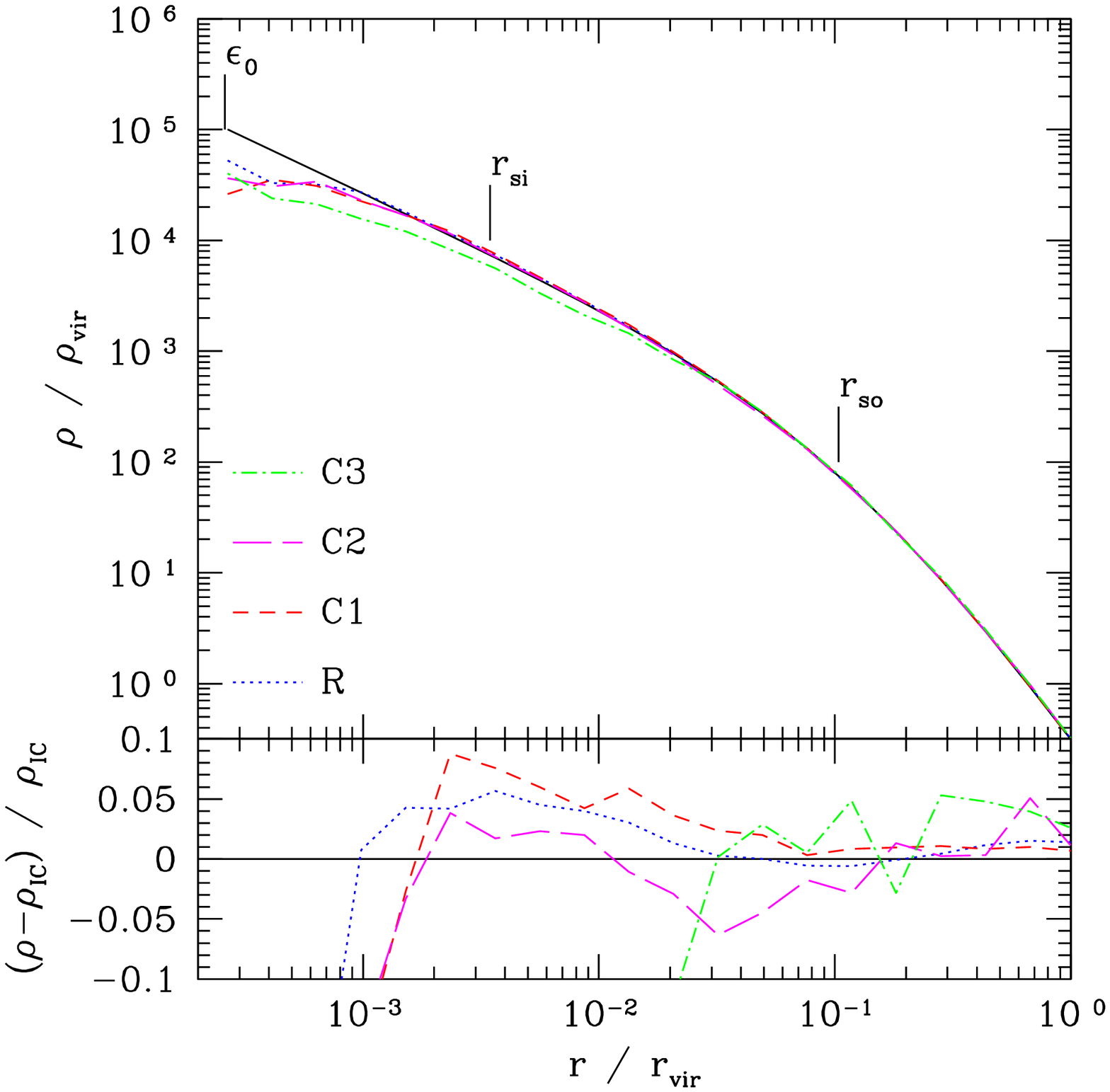} \\
		\caption{Stability plots for the more progressive models of series B (left) and C (right) after 10 Gyr. One can see that in general the profiles stay stable over 10 Gyr down to the relaxation scale and the perturbations are on a few per cent level - even for the reference single-mass model. Only for model C3 we pushed the refinement parameters too far and the structure is relatively strongly perturbed.}
		\label{fig:reftest}
\end{figure*}

\begin{table*}
	\caption{Summary of parameters for the different models with orbit dependent refinement. Also given are the theoretical and measured speed up factors for each model.}
	\label{tab:refmodels}
	\begin{tabular}{cccccccccc}
		\hline
		Model & A1 & A2 & A3 & B1 & B2 & B3 & C1 & C2 & C3\\
		\hline
		$\rso~[\rvir]$ &	
		1 &	
		1 &	
		1	&	
		$3.46 \times 10^{-1}$ &	
		$3.46 \times 10^{-1}$ &	
		$3.46 \times 10^{-1}$ &	
		$1.04 \times 10^{-1}$ & 
		$1.04 \times 10^{-1}$ &	
		$1.04 \times 10^{-1}$ \\
		$\rmor~[\rvir]$ & 
		$3.46 \times 10^{-2}$ &	
		$3.46 \times 10^{-2}$ & 
		$1.73 \times 10^{-2}$ & 
		$3.46 \times 10^{-2}$ &	
		$3.46 \times 10^{-2}$ & 
		$1.73 \times 10^{-2}$ & 
		$3.46 \times 10^{-2}$ &	
		$3.46 \times 10^{-2}$ & 
		$1.73 \times 10^{-2}$ \\ 		
		$\Nshell$ & 5 & 10 & 10 & 5 & 10 & 10 & 5 &	10 &10 \\
		$\Nsamp$ &
		$2.53 \times 10^{6}$ &
		$1.13 \times 10^{6}$ &
		$6.67 \times 10^{5}$ &
		$1.95 \times 10^{6}$ &
		$1.06 \times 10^{6}$ &
		$6.02 \times 10^{5}$ &
		$1.63 \times 10^{6}$ &
		$1.05 \times 10^{6}$ &
		$5.51 \times 10^{5}$ \\
		$\kappa$ & 0.612 & 1.22 & 1.22 & 0.753 & 1.51 & 1.51 & 1.02 & 2.04 & 2.04 \\
		$s_{\mathrm{th}}$ & 4.67 & 6.92 & 9.76 & 5.40 & 7.32 & 10.7 & 6.30 & 7.69 & 11.7 \\
		$s_{\mathrm{m}}$ & 5.75 & 8.82 & 12.3 & 6.61 & 8.43 & 13.3 & 7.51 & 8.72 & 12.9 \\			
		\hline
	\end{tabular}
	\medskip
\end{table*}

In the previous section, we presented models where the the heavier particles were allowed to penetrate the centre. Now, we focus on more general models with several particle species that have in addition the orbit dependent refinement described earlier. We performed a series of runs for a structure model with $\alpha = 1$, $\beta = 3$, $\gamma = 1$, and $\cvir = 10$. All models had $\rsi = 3.46 \times 10^{-3}~\rvir$, $N_0 = 10^4$, $\epsilon_0 = 2.59 \times 10^{-4}~\rvir$ in common. The other refinement parameters are given in Tab. \ref{tab:refmodels}. This results in a structure with an effective resolution within the virial radius of $\Nvir^\mathrm{eff} = 2.61 \times 10^7$ particles. For comparison, we also set up a single-mass model with the same resolution but without refinement as a reference (model R). Model R was sampled with a total of $\Nsamp = 3.54 \times 10^{7}$ particles including the cut-off region. We also kept $\Delta \RM = 2$ since we now have the parameter $\Nshell$ to control the overall mass range of the species and with that choice the mass contrast between neighbouring shells is kept to a minimum.

Each of these models was evolved for 10 Gyr in order to test for stability. The relaxation scale for such a halo is $\rrel(10~\Gyr,\Nvir^{\mathrm{eff}}) = 1.07 \times 10^{-3}~\rvir$. We present stability plots for the more progressive models of series B and C in Fig. \ref{fig:reftest}. We see that in general the profiles stay stable over 10 Gyr down to the relaxation scale and the perturbations are on a few per cent level - even for the reference single-mass model. Only for model C3 we pushed the refinement parameters too far (i.e. too small values for $\rso$ and $\rmor$) and the structure is relatively strongly perturbed. This was done deliberately in order to demonstrate that this technique has its limits and needs some careful parameter choice by the simulator that generates the initial conditions. We recommend in general to choose the values for $\rso$ and $\rmor$ not too small with respect to $\rsi$. This guarantees a larger protection zone around the high resolution centre and minimises the perturbations by the heavier particles.

The main advantage is again the computer run time gain. As we can see in Fig. \ref{fig:reftest}, many models which are stable down to a similar scale as the single-mass reference model R, are approximately an order of magnitude faster. In Tab. \ref{tab:refmodels}, we give for each model the values for $s_{\mathrm{th}}$ which we estimated from the discrete sampling of each structure (see equation (\ref{eq:sthd}))\footnote{It is in principle also possible to estimate $s_{\mathrm{th}}$ in the case with orbit dependent refinement from a similar continuous estimate like equation (\ref{eq:sthc}) for the case without orbit dependent refinement. This estimate corrects in addition for the mass fraction at a given radius $r$ that is splitted according to the splitting procedure described in section \ref{chap:odr}. Unfortunately, such an estimate is rather complicated to evaluate analytically and an estimate from the discrete sampling of the structure is much more practical.} and $s_{\mathrm{m}} = T_0/T$ where $T_0$ is the computer run time for the single-mass reference model R and $T$ is the computer run time of the specific multimass model. We also see that a more progressive choice of $\kappa$ as for example between model B1 and B2 can lead to a 20\% speed up. Of course, we ran all models under the same conditions, i.e the same number of CPUs on the same super-computer.

\begin{figure*}
	\centering
		\includegraphics[width=0.495\textwidth]{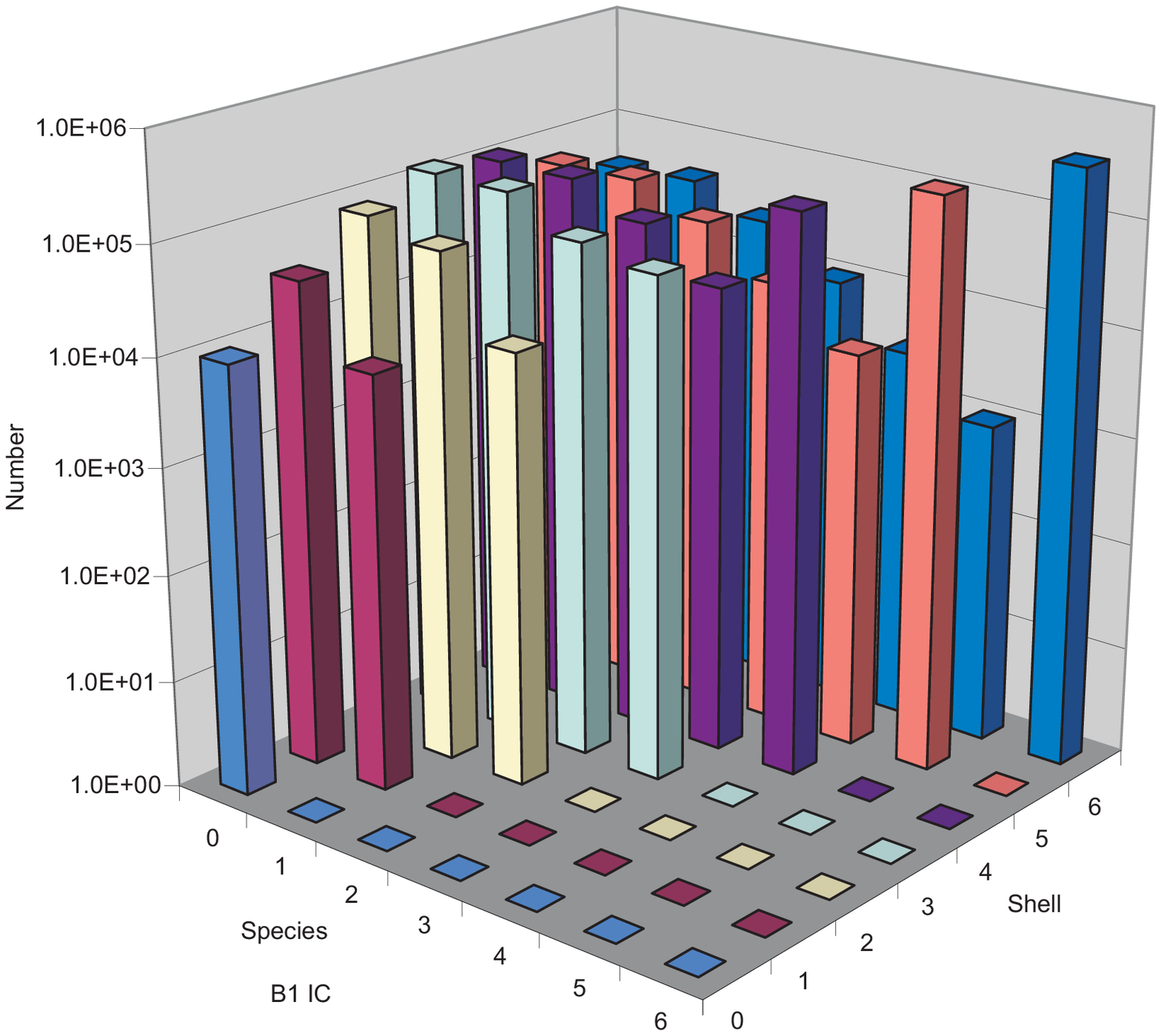} \hfill
		\includegraphics[width=0.495\textwidth]{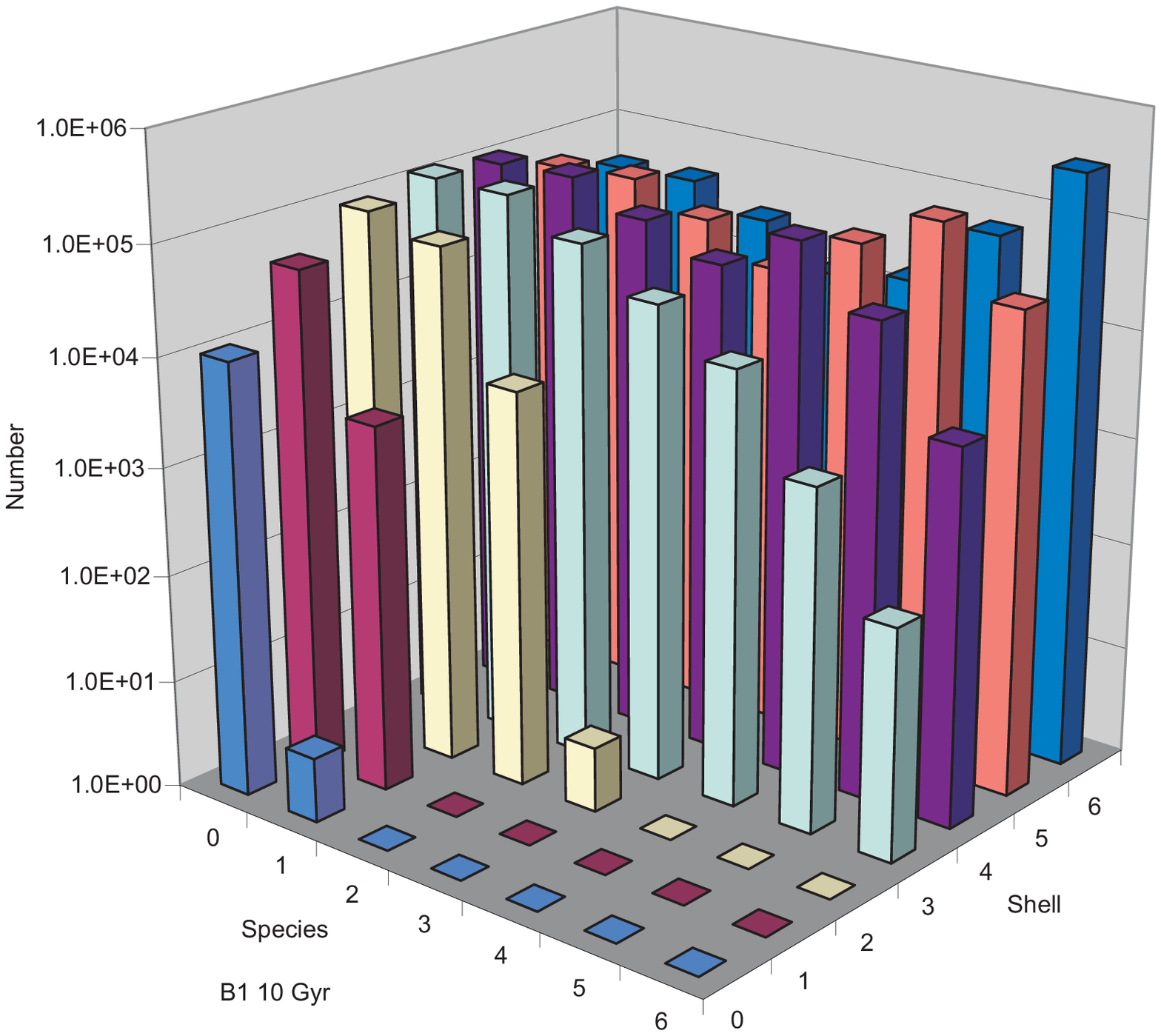} \\
		\includegraphics[width=0.495\textwidth]{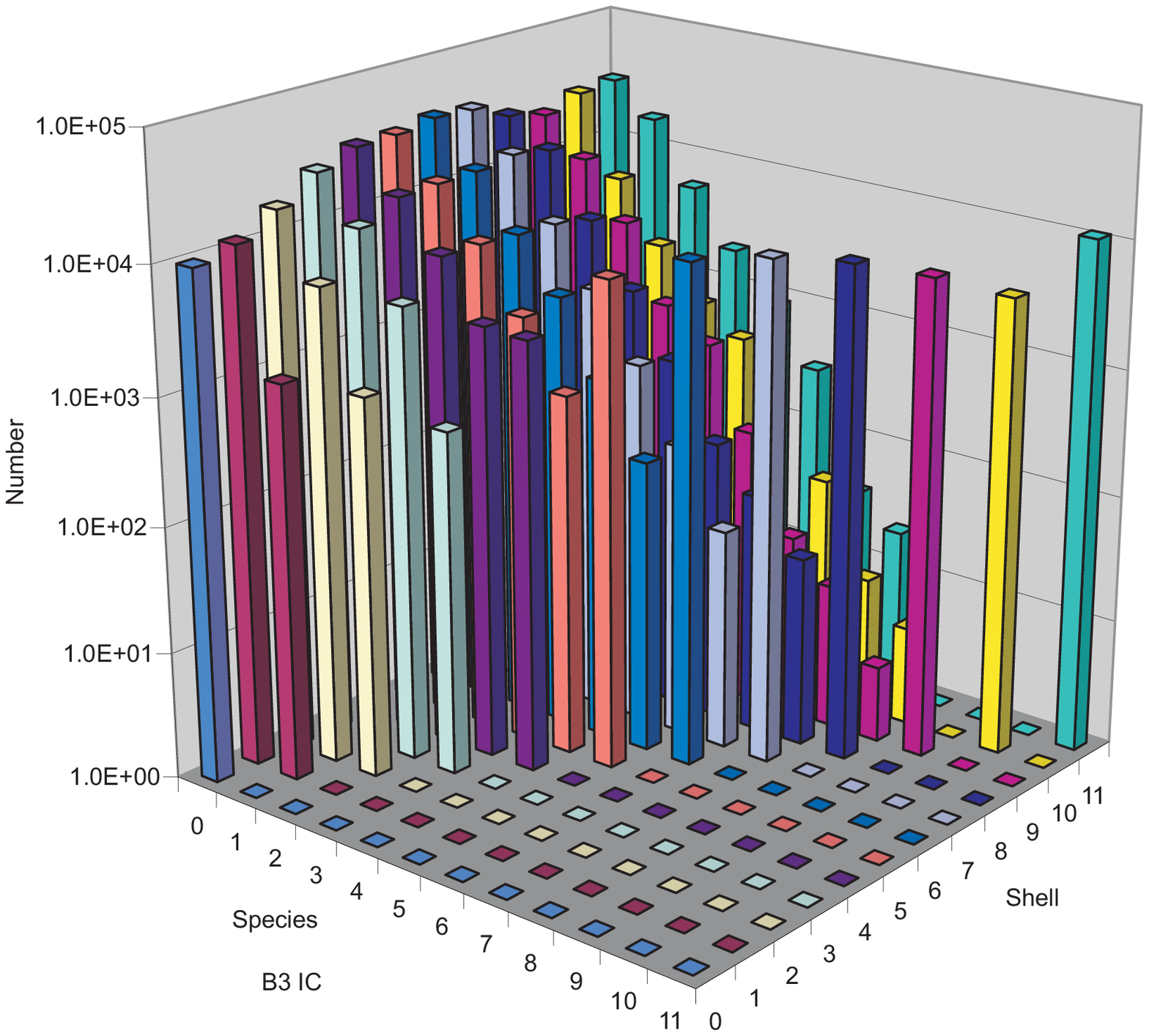} \hfill
		\includegraphics[width=0.495\textwidth]{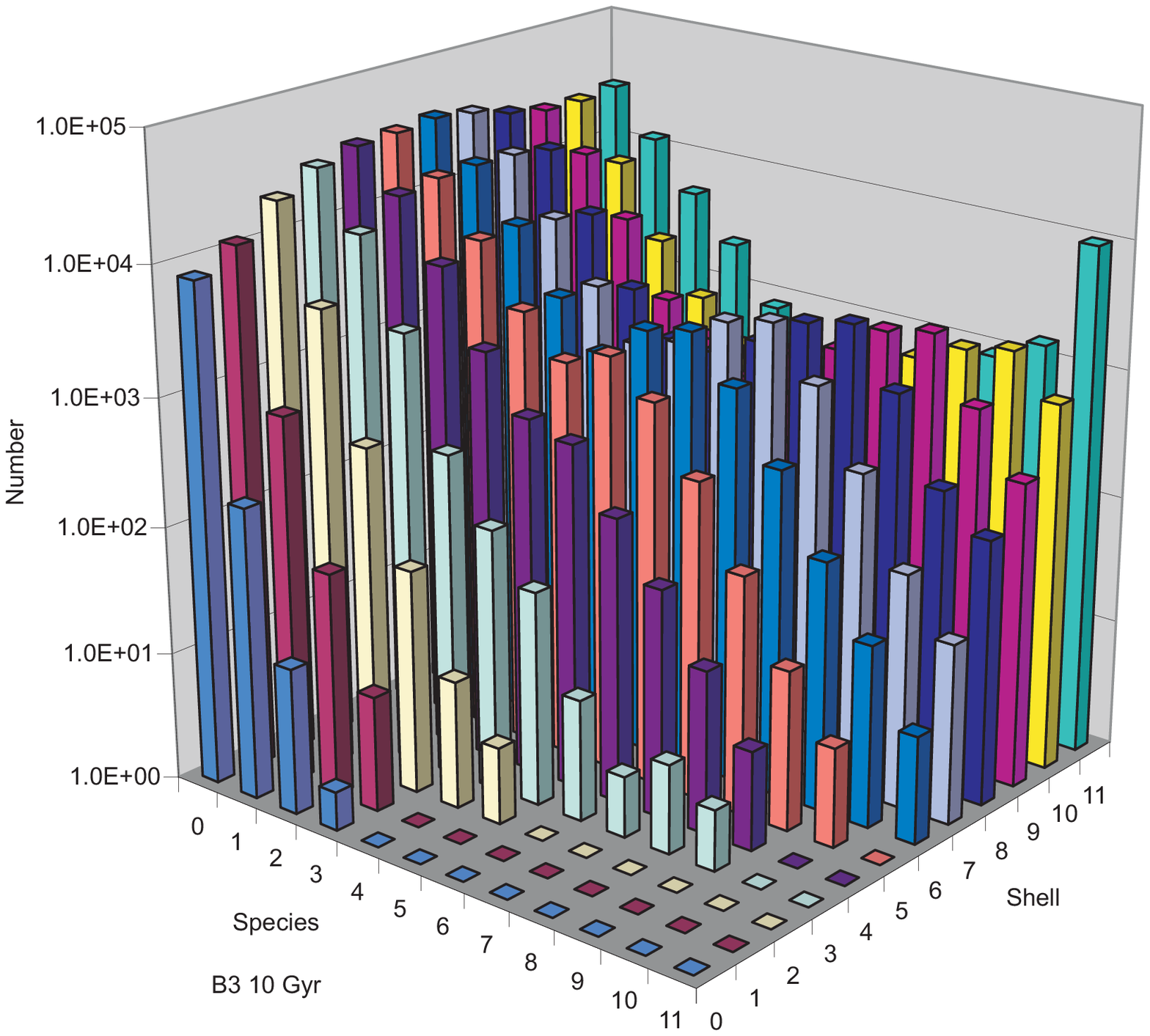} \\
		\caption{Histograms of the particle distribution for the models B1 (top) and B3 (bottom) for the initial conditions (left) and after 10 Gyr (right). We see that even after 10 Gyr only a few particles with masses heavier than $m_0$ are in the innermost shell and that none of the very heavy particles are closer than $\rmor$ in both cases.}
		\label{fig:mixing}
\end{figure*}

We now illustrate the mixing of the different particle species in more detail on the basis of model B1 and B3. For B1 we chose $\Nshell = 5$ and $\rmor = 3.46 \times 10^{-2}~\rvir$ whereas for B3 we chose some more progressive values of $\Nshell = 10$ and $\rmor = 1.73 \times 10^{-2}~\rvir$. The model B3 is then sampled by less than a third of the particles needed for B1 and the computer run time for model B3 is approximately half the value of B1. If we say we trust the dynamics down to the scale where the relative perturbations $(\rho-\rho_{\mathrm{IC}})/\rho_{\mathrm{IC}}$ are not larger than 10\%, then we lose approximately a factor of two in resolution for model B3 compared with B1 (see also Fig. \ref{fig:reftest}).

In Fig. \ref{fig:mixing}, we plot histograms of the particle distribution for both models, B1 (top) and B3 (bottom), for the initial conditions (left) and after 10 Gyr (right). The $\Nshell$ shells with index $1\ldots\Nshell$ are defined by the values of $\rsi$ and $\rso$. Between these two values, the shells are equally spaced logarithmically. The shell with index 0 contains everything smaller than $\rsi$ and the shell with index $\Nshell+1$ includes everything larger than $\rso$. The ranges for the mass species are defined in a similar way: the mass species with index $i$ contains the range $m_{i-1} < m \leq m_i$ where $m_i = m_0 (\Delta \RM)^i$ and $m_{-1} \equiv 0$. We see that even after 10 Gyr only a few particles with masses heavier than $m_0$ are in the innermost shell. For example in model B1, only four particles of mass $m_1$ are in the innermost shell after 10 Gyr. A few more particles from neighbouring shells populate the innermost shell in model B3, which has a less conservative choice of refinement parameters. For model B1 we chose $\rmor = 10~\kpc$ which lies in the middle of shell 3 and for B3 we chose $\rmor = 5~\kpc$ which lies in the middle of shell 4. We see that none of the very heavy particles are closer than $\rmor$ in both cases. This tells us that the orbit dependent refinement procedure works very well and we do not get scattering of heavy particles on orbits that go through the centre.

\subsection{A structure with $1.68 \times 10^9$ particles}

\begin{figure}
	\centering
		\includegraphics[width=\columnwidth]{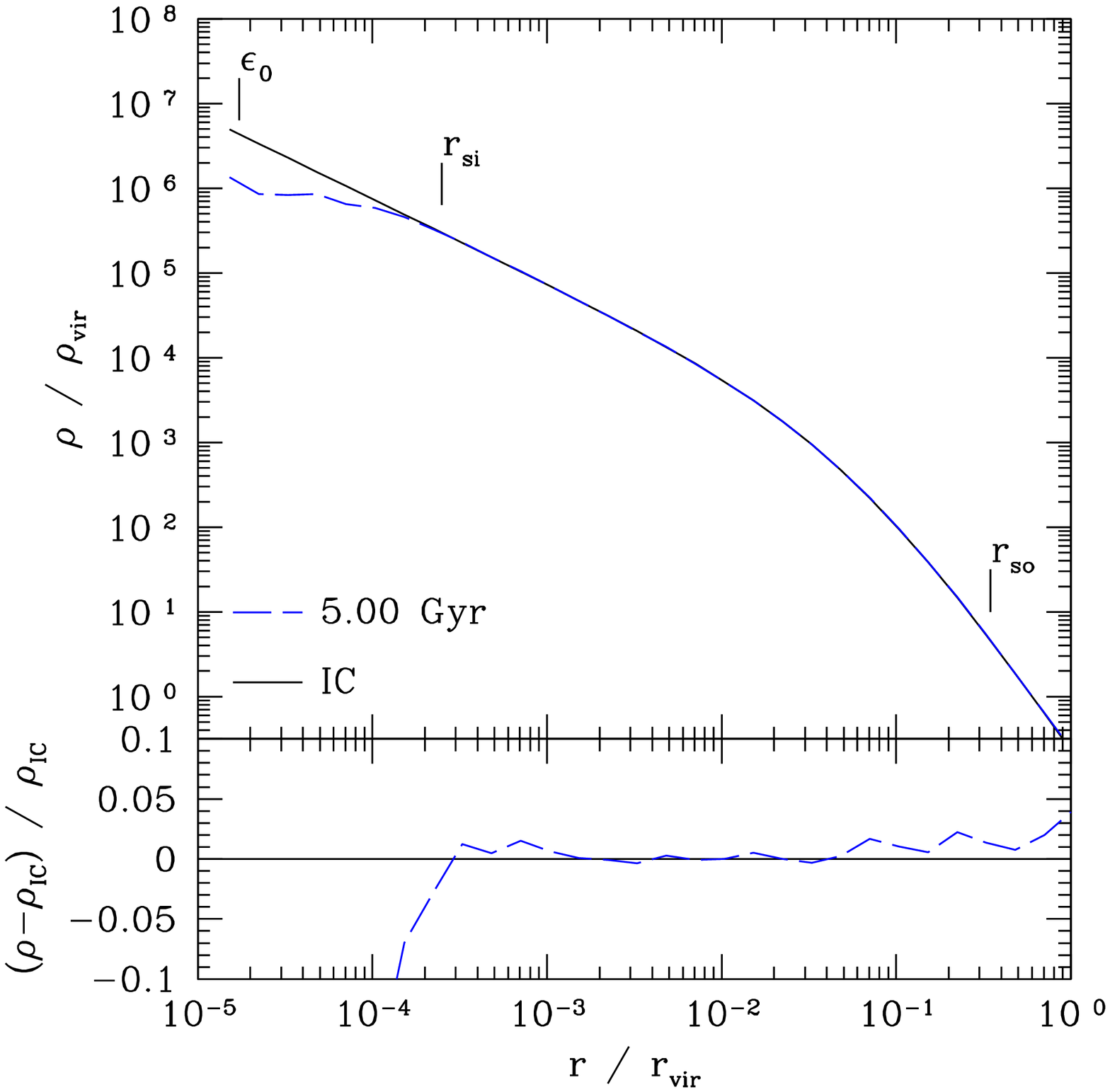}
		\caption{Density profile of an equilibrium structure with $\Nvir^{\mathrm{eff}} = 1.68 \times 10^9$ particles after 5 Gyr.}
		\label{fig:bigone}
\end{figure}

As a last test we present a structure with an effective resolution of $\Nvir^{\mathrm{eff}} = 1.68 \times 10^9$ particles. We chose $\alpha = 1$, $\beta = 3$, $\gamma = 1$, $\cvir = 20$, $N_0 = 10^4$, $\epsilon_0 = 1.73 \times 10^{-5}~\rvir$, $\rsi = 5 \times 10^{-3}~\rs$, $\rso = 3.46 \times 10^{-1}~\rvir$, $\Nshell = 12$, $\kappa = 1.15$ and $\rmor = 6.91 \times 10^{-3}~\rvir$. The theoretical speed up factore for this structure is $s_{\mathrm{th}} = 28.6$ which we again estimated from the sampled structure via equation (\ref{eq:sthd}). We evolved this structure for 5 Gyr and the density plot can be seen in Fig. \ref{fig:bigone}. The relaxation scale is $\rrel(5~\Gyr,\Nvir^{\mathrm{eff}}) = 1.31 \times 10^{-4}~\rvir$ and we see that this structure is stable down to that scale.

\section{Preservation of cusp slope} \label{chap:preservation}

\citet{2005MNRAS.360..892D} showed analytically that in a merger of self-gravitating cusps with different central slopes the merger remnant has always the slope of the steepest progenitor. In other words the steepest cusp slope is preserved. 

Collisionless mergers of dark matter halos were already studied in earlier work \citep[e.g.][]{1999IAUS..186..137B,2004MNRAS.349.1117B,2004MNRAS.354..522M,2006RMxAA..42...41A,2006ApJ...641..647K} but none of these studies had the resolution of the simulations presented here. We do not find any discrepancies between this work and the earlier studies lending support to the fact that the multimass technique works under the extreme dynamics of a dark matter halo merger event.

We initially set-up two dark matter halo models with different central slope in isolation as in section \ref{chap:tests}. Both profiles were from the $\alpha\beta\gamma$-model family and had the following general specifications: $\alpha = 1$, $\beta = 3$, $\cvir = 10$, $\Mvir = 1.43 \times 10^{12}~\Mo$ and $\rvir = 289~\kpc$. One structure had a central slope of $\gamma = 0$, the other had $\gamma = 1$. For both models we used $\Delta \RM = 2$ and the other parameters are given in Tab. \ref{tab:icmerger}. The resolution is clearly limited by the centrally flat ($\gamma = 0$) model since here many more particles are needed to resolve a given scale compared to steeper profiles. To resolve the same scale with the steeper profile then fewer particles would be needed, but since we would not like to have too high mass ratios between the two high resolution species in the centre we increased the number of particles in the $\gamma = 1$ structure. The theoretical speed up factor is $s_{\mathrm{th}} = 12.0$ for the model with $\gamma = 0$ and $s_{\mathrm{th}} = 7.39$ for the model with $\gamma = 1$ which was again estimated from the discrete sampling of the structures.

\begin{table}
	\caption{Summary of parameters for the two models used for the mergers.}
	\label{tab:icmerger}
	\begin{tabular}{ccc}
		\hline
		$\gamma$ & 0 & 1 \\
		\hline
		$N_0$ & $1 \times 10^4$ & $4 \times 10^4$ \\
		$\epsilon_0~[\rvir]$ & 
		$3.46 \times 10^{-4}$ & 
		$3.46 \times 10^{-4}$ \\
		$\Nvir^{\mathrm{eff}}$ &
		$1.14 \times 10^{8}$ &
		$1.04 \times 10^{8}$ \\
		$\Nsamp$ &
		$5.67 \times 10^{6}$ &
		$4.29 \times 10^{6}$ \\		
		$\rsi~[\rvir]$ & 
		$6.91 \times 10^{-3}$ & 
		$3.46 \times 10^{-3}$ \\
		$\rso~[\rvir]$ & 
		$3.46 \times 10^{-1}$ & 
		$3.46 \times 10^{-1}$ \\		
		$\rmor~[\rvir]$ & 
		$6.91 \times 10^{-2}$ & 
		$3.46 \times 10^{-2}$ \\				
		$\Nshell$ & 10 & 10\\			
		$\kappa$ & 1.77 & 1.51 \\
		$\rrel(10~\Gyr)~[\rvir]$ &
		$1.19 \times 10^{-3}$ &
		$6.26 \times 10^{-4}$ \\
		\hline
	\end{tabular}
	\medskip
\end{table}

\begin{figure*}
	\centering
		\includegraphics[width=0.495\textwidth]{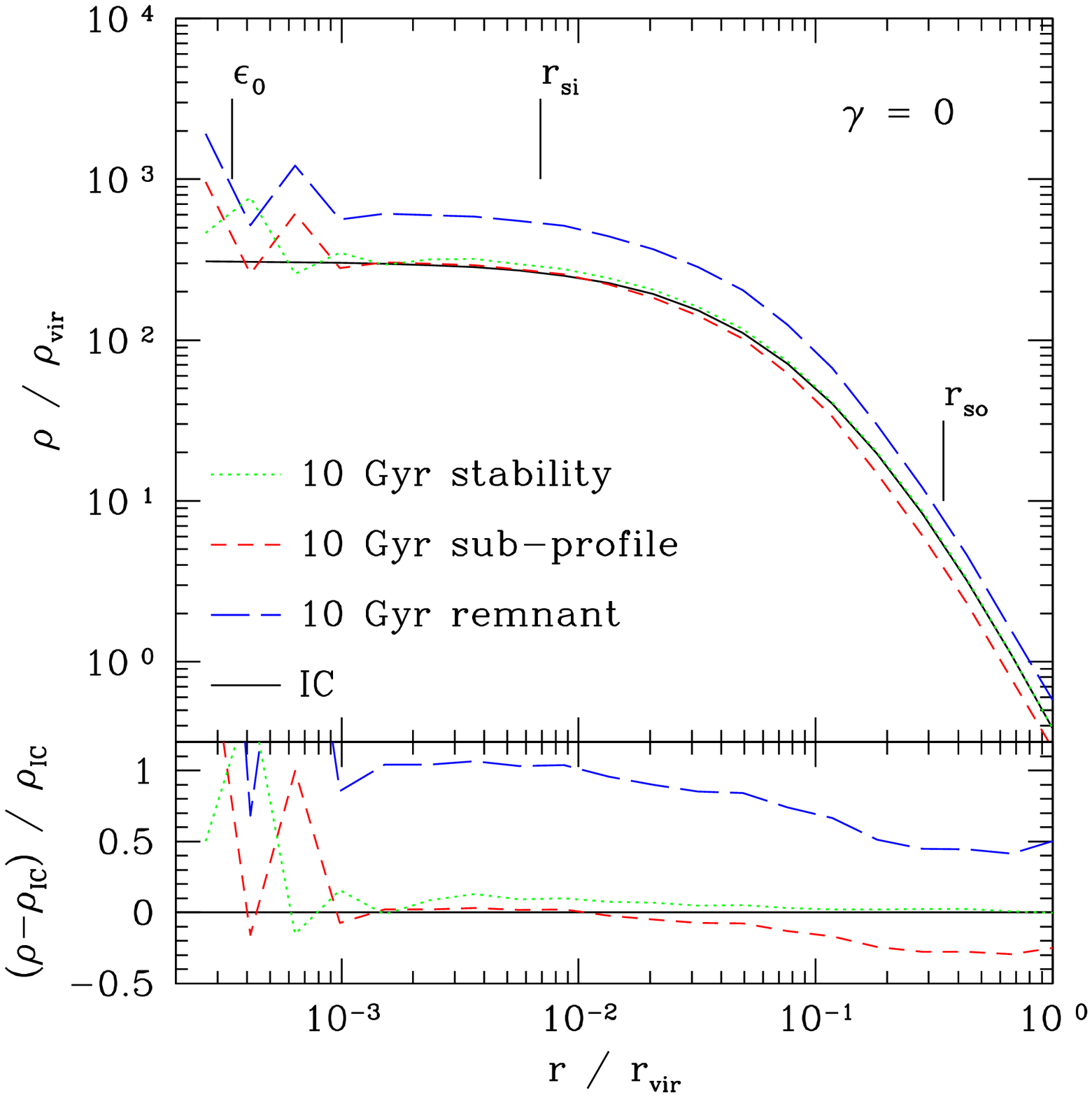} \hfill
		\includegraphics[width=0.495\textwidth]{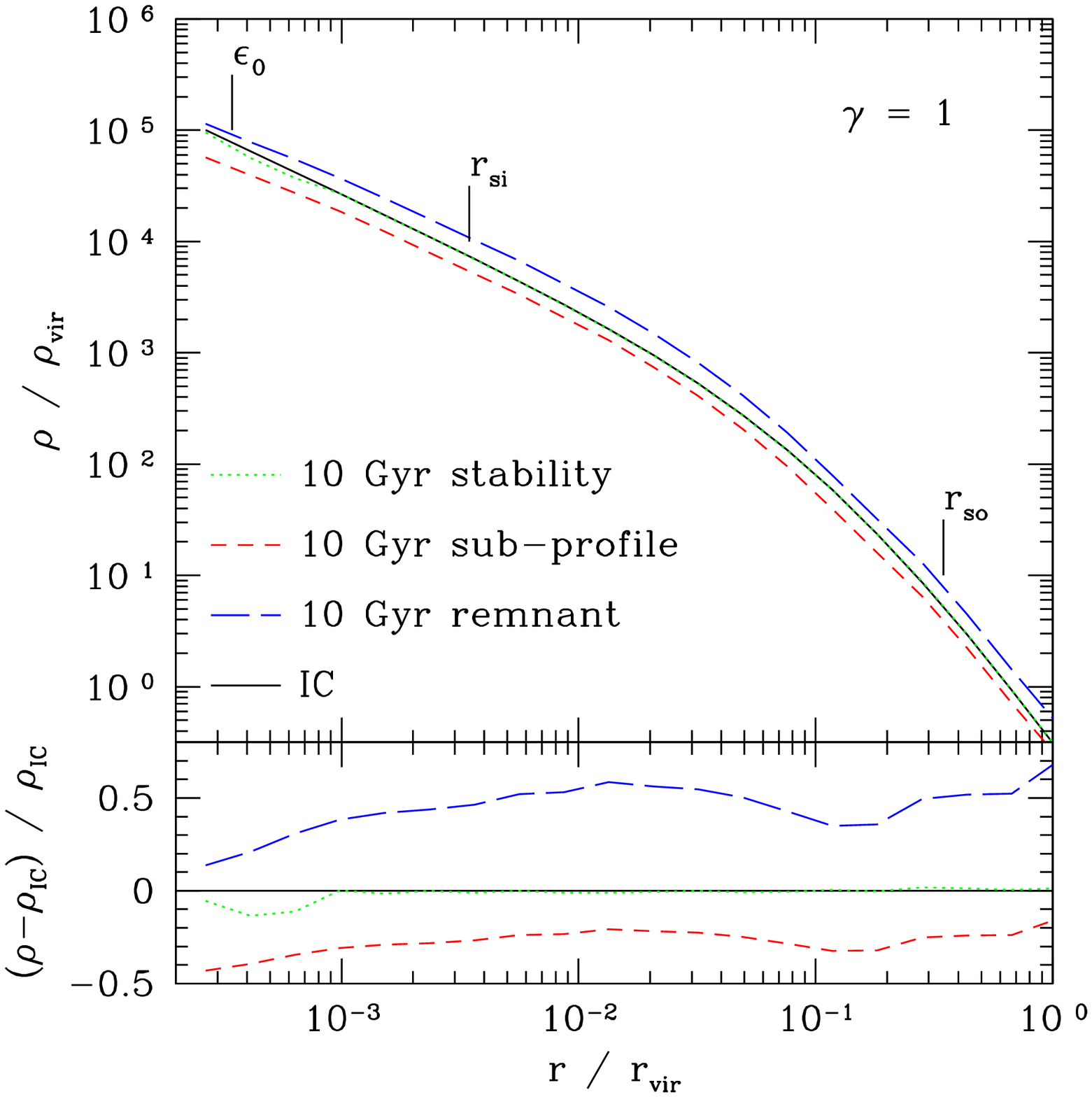} \\
		\includegraphics[width=0.495\textwidth]{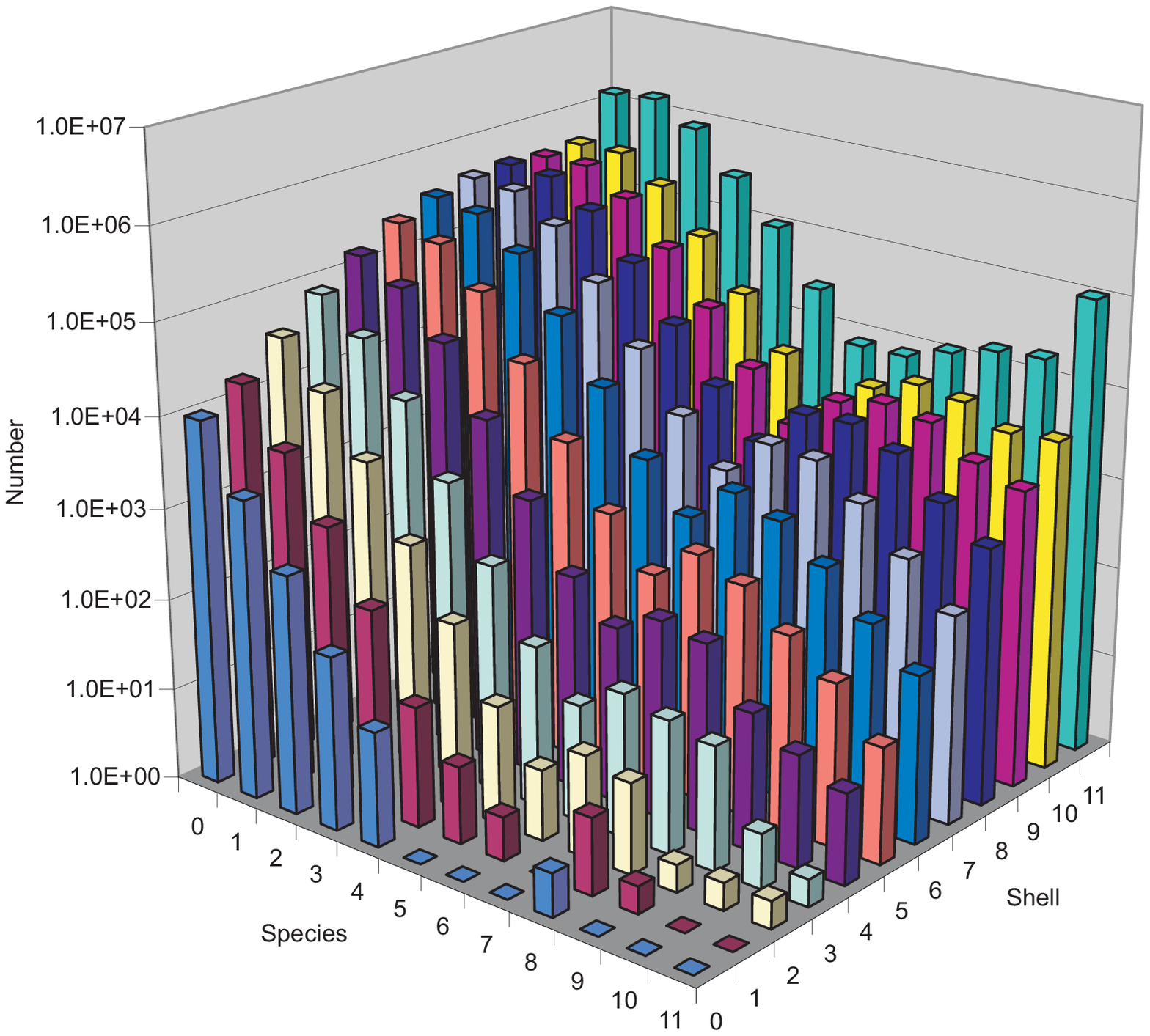} \hfill
		\includegraphics[width=0.495\textwidth]{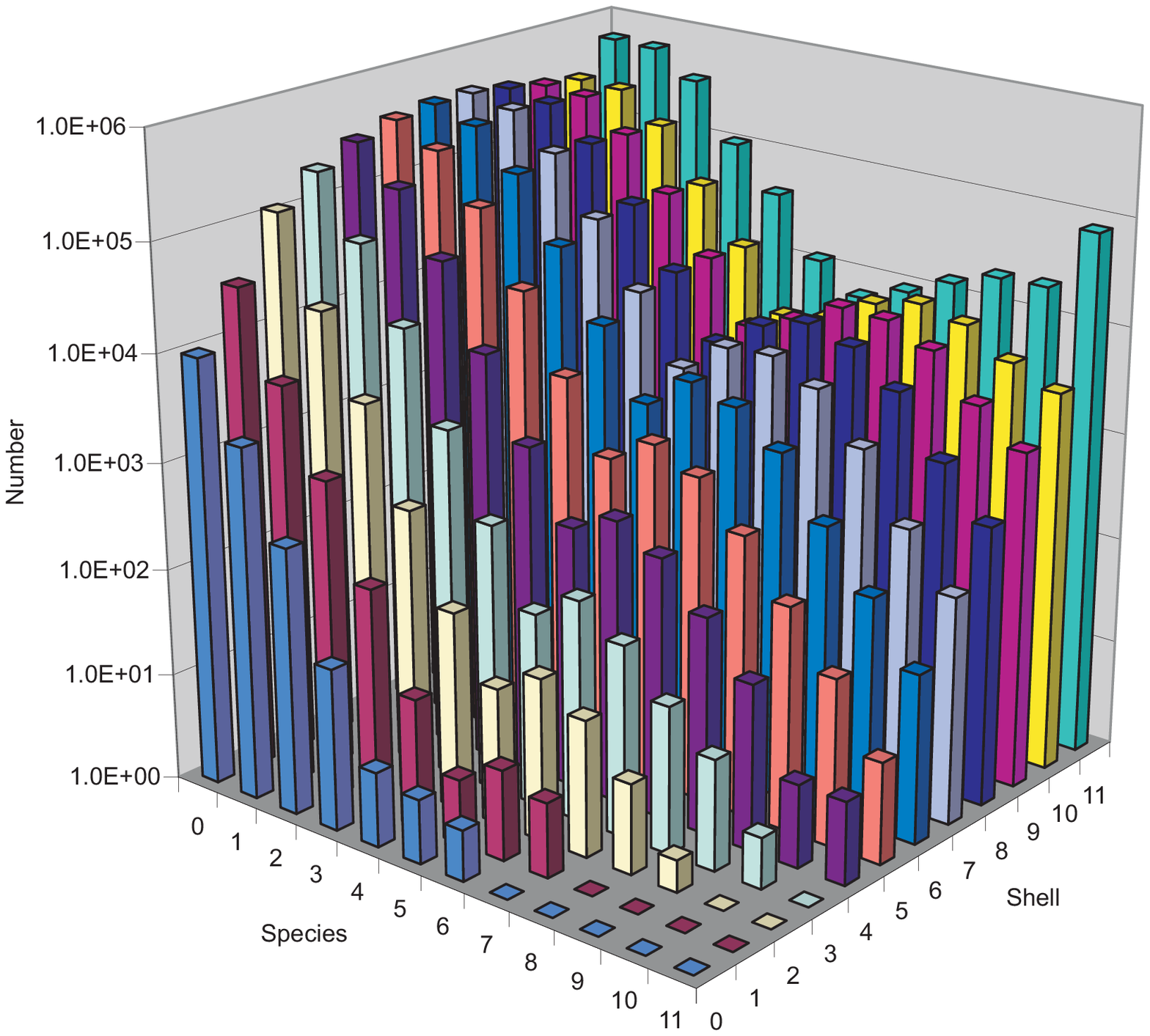} \\
		\caption{Equal slope mergers, left column: $\gamma = 0$ and right column: $\gamma = 1$. Top row: The profiles are stable over 10 Gyr and the central slope of the structure is conserved in both cases. Interestingly, we find that in the $\gamma = 0$ case the central density approximately doubles whereas in the $\gamma = 1$ case the central density in the merger remnant only increases by approximately 50\%. This can also be seen in the average sub-profile of the two individual structures. Bottom row: histograms of the particle distribution after 10 Gyr. We can again see that only a few heavy particles are in the innermost shell.}
		\label{fig:merger1}
\end{figure*}

We evolved both halos in isolation for 10 Gyr in order to test the stability again. In total three mergers were performed: a cusp-cusp, a core-core and a cusp-core merger where cusp means $\gamma = 1$ and core is equivalent to the $\gamma = 0$. The following merger set-up was used for all three cases. We placed the two individual haloes $600~\kpc \approx 2~\rvir$ apart and the two haloes had an initial relative radial velocity of $v_{\mathrm{rad}} = 150~\kpc~\Gyr^{-1}$ and a relative tangential velocity of $v_{\mathrm{tan}} = 50~\kpc~\Gyr^{-1}$. Assuming the two haloes were point masses, this set-up corresponds to an eccentricity $e \approx 0.95$ of the orbit consistent with values in cosmological $N$-body simulations \citep{2006A&A...445..403K}. With this set-up the merger time needed by the two haloes to merge completely was approximately 7 Gyr. We let all three runs evolve to 10 Gyr so that the merger remnant has time to relax.

Fig. \ref{fig:merger1} the top row shows the stability of both models and the equal profile mergers after 10 Gyr. Down to a few softening lengths of the lightest particles no significant deviations can be seen and the merger remnant has the same central profile as the two progenitors. We normalise the plots by the values $\rvir$ and $\rhovir$ of the progenitor haloes. Interestingly, we find that in the $\gamma = 0$ case the central density approximately doubles whereas in the $\gamma = 1$ case the central density in the merger remnant only increases by approximately 50\%. This can also be seen in the average sub-profile of the two individual structures which stays constant in the $\gamma = 0$ case whereas it decreases by approximately 25\% in the cusp case with $\gamma = 1$. The case with $\gamma = 1$ is similar to a major merger in a cosmological simulation where one finds the same effect \citep{2006astro.ph..9615F}. This may suggest that the central region of flat structures are better protected and get less energy input from the merger. If this is a general feature of cored profiles and to what degree the merger configuration plays a role is unclear and needs to be investigated further. The bottom row shows histograms of the particle distribution after 10 Gyr. The particle species were defined in the same way as for Fig. \ref{fig:mixing}. Even under the violent dynamcis of a merger, the contamination of the innermost shell with heavier particles is relatively small and hence does not affect the central slopes of the density profiles.

\begin{figure}
	\centering
		\includegraphics[width=\columnwidth]{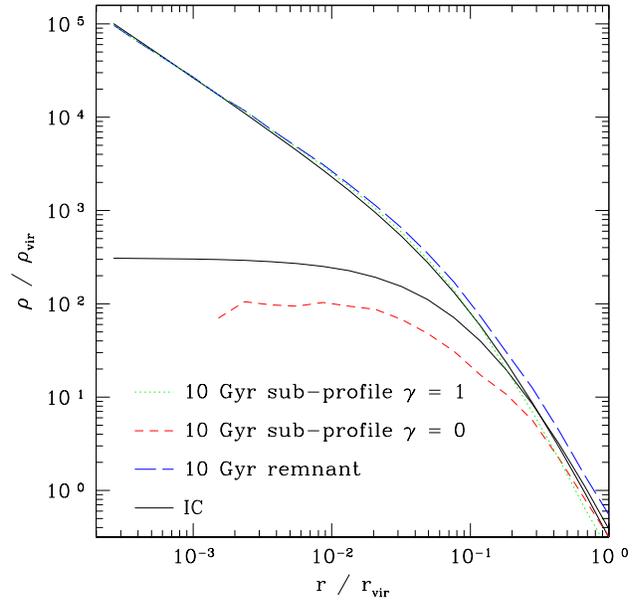}
		\caption{Cusp-core merger after 10 Gyr. The central cusp is preserved. As a reference we also pot the initial profiles of two individual structures.}
		\label{fig:merger2}
\end{figure}

In Fig. \ref{fig:merger2}, we present the profile of the cusp-core merger after 10 Gyr. The central slope of the steepest progenitor is perfectly preserved and the cored progenitor only contributes significantly in the outer region of the density profile. If we look at the sub-profile of the particles that belonged initially to the cored halo, we now see that in this case the cored structure gets much more energy input from the merger and its central density is lowered approximately by a factor of three through the merger process.

We can therefore confirm the earlier findings that core-core mergers lead to a cored merger remnant while cups-cusp mergers lead to a cuspy merger remnant with high resolution multimass $N$-body simulations. In cusp-core mergers the merger remnant has a final profile corresponding to the steepest progenitor which is in excellent agreement with the theoretical predictions by \citet{2005MNRAS.360..892D}.

\section{Conclusions} \label{chap:conclusions}

The multimass technique for modelling haloes is a simple method to perform high resolution $N$-body simulations. An early version of this technique without orbit dependent refinement has already been used successfully in several applications which also include cosmological structure formation simulations \citep{2005MNRAS.364..665D,2006MNRAS.368.1073G}. With a careful choice of parameters it is possible to gain over an order of magnitude in computer run time for a given resolution scale.

As an application of this technique we confirm the earlier findings that core-core mergers lead to a cored merger remnant while cups-cusp mergers lead to a cuspy merger remnant with high resolution multimass $N$-body simulations. In cusp-core merges, the merger remnant has a final profile corresponding to the steepest progenitor which is in excellent agreement with the theoretical predictions. We find that in the core-core case the central density approximately doubles whereas in the cusp-cusp case the central density in the merger remnant only increases by approximately 50\%. This may suggest that the central region of flat structures are better protected and get less energy input from the merger.

A software tool called \textsc{halogen} (HALO GENerator) for generating multimass initial conditions is available from the author upon request.

\section*{Acknowledgment}

It is a pleasure to thank J\"urg Diemand for many discussions and helpful comments. All the simulations were carried out at the zBox2\footnote{http://www.zbox2.org} supercomputer at University Zurich and the Pleiades\footnote{http://pleione.ucsc.edu/pleiades} supercomputer at University of California Santa Cruz. M.Z. was supported by a fellowship from Swiss National Science Foundation. P.M. acknowledges support from NASA grant NNG04GK85G and NSF grant 0521566.

\bibliography{RDB_S}

\begin{thebibliography}{}

\bibitem[\protect\citeauthoryear{{Aceves} \& {Vel{\'a}zquez}}{{Aceves} \&
  {Vel{\'a}zquez}}{2006}]{2006RMxAA..42...41A}
{Aceves} H.,  {Vel{\'a}zquez} H.,  2006, RMxAA, 42, 41

\bibitem[\protect\citeauthoryear{{Adams} \& {Laughlin}}{{Adams} \&
  {Laughlin}}{1997}]{1997RvMP...69..337A}
{Adams} F.~C.,  {Laughlin} G.,  1997, Rev. Mod. Phys., 69, 337

\bibitem[\protect\citeauthoryear{{Barnes}}{{Barnes}}{1999}]{1999IAUS..186..137%
B}
{Barnes} J.~E.,  1999, in {Barnes} J.~E.,  {Sanders} D.~B.,  eds, Galaxy
  Interactions at Low and High Redshift Vol.~186 of IAU Symposium, {D}ynamics
  of {M}ergers \& {R}emnants.
pp 137--+

\bibitem[\protect\citeauthoryear{{Binney} \& {Tremaine}}{{Binney} \&
  {Tremaine}}{1987}]{1987gady.book.....B}
{Binney} J.,  {Tremaine} S.,  1987, {G}alactic dynamics, 3. edn.
Princeton, NJ, Princeton University Press

\bibitem[\protect\citeauthoryear{{Boylan-Kolchin} \& {Ma}}{{Boylan-Kolchin} \&
  {Ma}}{2004}]{2004MNRAS.349.1117B}
{Boylan-Kolchin} M.,  {Ma} C.-P.,  2004, MNRAS, 349, 1117

\bibitem[\protect\citeauthoryear{{Chenciner} \& {Montgomery}}{{Chenciner} \&
  {Montgomery}}{2000}]{2000AM....152..881C}
{Chenciner} A.,  {Montgomery} R.,  2000, Ann. Math., 152, 881

\bibitem[\protect\citeauthoryear{{Corless}, {Gonnet}, {Hare}, {Jeffrey} \&
  {Knuth}}{{Corless} et~al.}{1996}]{1996ACM.....5..329C}
{Corless} R.~M.,  {Gonnet} G.~H.,  {Hare} D.~E.~G.,  {Jeffrey} D.~J.,
  {Knuth} D.~E.,  1996, Adv. Comput. Math., 5, 329

\bibitem[\protect\citeauthoryear{{Dehnen}}{{Dehnen}}{1993}]{1993MNRAS.265..250%
D}
{Dehnen} W.,  1993, MNRAS, 265, 250

\bibitem[\protect\citeauthoryear{{Dehnen}}{{Dehnen}}{2005}]{2005MNRAS.360..892%
D}
{Dehnen} W.,  2005, MNRAS, 360, 892

\bibitem[\protect\citeauthoryear{{Diemand}, {Kuhlen} \& {Madau}}{{Diemand}
  et~al.}{2007a}]{2007ApJ...657..262D}
{Diemand} J.,  {Kuhlen} M.,    {Madau} P.,  2007a, ApJ, 657, 262

\bibitem[\protect\citeauthoryear{{Diemand}, {Kuhlen} \& {Madau}}{{Diemand}
  et~al.}{2007b}]{2007ApJ...667..859D}
{Diemand} J.,  {Kuhlen} M.,    {Madau} P.,  2007b, ApJ, 667, 859

\bibitem[\protect\citeauthoryear{{Diemand}, {Moore}, {Stadel} \&
  {Kazantzidis}}{{Diemand} et~al.}{2004}]{2004MNRAS.348..977D}
{Diemand} J.,  {Moore} B.,  {Stadel} J.,    {Kazantzidis} S.,  2004, MNRAS,
  348, 977

\bibitem[\protect\citeauthoryear{{Diemand}, {Zemp}, {Moore}, {Stadel} \&
  {Carollo}}{{Diemand} et~al.}{2005}]{2005MNRAS.364..665D}
{Diemand} J.,  {Zemp} M.,  {Moore} B.,  {Stadel} J.,    {Carollo} C.~M.,  2005,
  MNRAS, 364, 665

\bibitem[\protect\citeauthoryear{{Dyson}}{{Dyson}}{1979}]{1979RvMP...51..447D}
{Dyson} F.~J.,  1979, Rev. Mod. Phys., 51, 447

\bibitem[\protect\citeauthoryear{{Eke}, {Navarro} \& {Frenk}}{{Eke}
  et~al.}{1998}]{1998ApJ...503..569E}
{Eke} V.~R.,  {Navarro} J.~F.,    {Frenk} C.~S.,  1998, ApJ, 503, 569

\bibitem[\protect\citeauthoryear{{Faltenbacher}, {Gottloeber} \&
  {Mathews}}{{Faltenbacher} et~al.}{2006}]{2006astro.ph..9615F}
{Faltenbacher} A.,  {Gottloeber} S.,    {Mathews} W.~G.,  2006, preprint
  (astro-ph/0609615)

\bibitem[\protect\citeauthoryear{{Goerdt}, {Moore}, {Read}, {Stadel} \&
  {Zemp}}{{Goerdt} et~al.}{2006}]{2006MNRAS.368.1073G}
{Goerdt} T.,  {Moore} B.,  {Read} J.~I.,  {Stadel} J.,    {Zemp} M.,  2006,
  MNRAS, 368, 1073

\bibitem[\protect\citeauthoryear{{Hernquist}}{{Hernquist}}{1990}]{1990ApJ...35%
6..359H}
{Hernquist} L.,  1990, ApJ, 356, 359

\bibitem[\protect\citeauthoryear{{Islam}}{{Islam}}{1977}]{1977QJRAS..18....3I}
{Islam} J.~N.,  1977, QJRAS, 18, 3

\bibitem[\protect\citeauthoryear{{Kazantzidis}, {Magorrian} \&
  {Moore}}{{Kazantzidis} et~al.}{2004}]{2004ApJ...601...37K}
{Kazantzidis} S.,  {Magorrian} J.,    {Moore} B.,  2004, ApJ, 601, 37

\bibitem[\protect\citeauthoryear{{Kazantzidis}, {Zentner} \&
  {Kravtsov}}{{Kazantzidis} et~al.}{2006}]{2006ApJ...641..647K}
{Kazantzidis} S.,  {Zentner} A.~R.,    {Kravtsov} A.~V.,  2006, ApJ, 641, 647

\bibitem[\protect\citeauthoryear{{Khochfar} \& {Burkert}}{{Khochfar} \&
  {Burkert}}{2006}]{2006A&A...445..403K}
{Khochfar} S.,  {Burkert} A.,  2006, A\&A, 445, 403

\bibitem[\protect\citeauthoryear{{Kuijken} \& {Dubinski}}{{Kuijken} \&
  {Dubinski}}{1994}]{1994MNRAS.269...13K}
{Kuijken} K.,  {Dubinski} J.,  1994, MNRAS, 269, 13

\bibitem[\protect\citeauthoryear{{Lambert}}{{Lambert}}{1758}]{1758AH......3..1%
28L}
{Lambert} J.~H.,  1758, Acta Helvetica,
  physico-mathematico-anatomico-botanico-medica, 3, 128

\bibitem[\protect\citeauthoryear{{Leeuwin}, {Combes} \& {Binney}}{{Leeuwin}
  et~al.}{1993}]{1993MNRAS.262.1013L}
{Leeuwin} F.,  {Combes} F.,    {Binney} J.,  1993, MNRAS, 262, 1013

\bibitem[\protect\citeauthoryear{{Moore}, {Governato}, {Quinn}, {Stadel} \&
  {Lake}}{{Moore} et~al.}{1998}]{1998ApJ...499L...5M}
{Moore} B.,  {Governato} F.,  {Quinn} T.,  {Stadel} J.,    {Lake} G.,  1998,
  ApJ Lett., 499, L5+

\bibitem[\protect\citeauthoryear{{Moore}, {Kazantzidis}, {Diemand} \&
  {Stadel}}{{Moore} et~al.}{2004}]{2004MNRAS.354..522M}
{Moore} B.,  {Kazantzidis} S.,  {Diemand} J.,    {Stadel} J.,  2004, MNRAS,
  354, 522

\bibitem[\protect\citeauthoryear{{Moore}}{{Moore}}{1993}]{1993PhRvL..70.3675M}
{Moore} C.,  1993, Phys. Rev. Lett., 70, 3675

\bibitem[\protect\citeauthoryear{{Navarro}, {Frenk} \& {White}}{{Navarro}
  et~al.}{1996}]{1996ApJ...462..563N}
{Navarro} J.~F.,  {Frenk} C.~S.,    {White} S.~D.~M.,  1996, ApJ, 462, 563

\bibitem[\protect\citeauthoryear{{Power}, {Navarro}, {Jenkins}, {Frenk},
  {White}, {Springel}, {Stadel} \& {Quinn}}{{Power}
  et~al.}{2003}]{2003MNRAS.338...14P}
{Power} C.,  {Navarro} J.~F.,  {Jenkins} A.,  {Frenk} C.~S.,  {White} S.~D.~M.,
   {Springel} V.,  {Stadel} J.,    {Quinn} T.,  2003, MNRAS, 338, 14

\bibitem[\protect\citeauthoryear{{Prada}, {Klypin}, {Simonneau},
  {Betancort-Rijo}, {Patiri}, {Gottl{\"o}ber} \& {Sanchez-Conde}}{{Prada}
  et~al.}{2006}]{2006ApJ...645.1001P}
{Prada} F.,  {Klypin} A.~A.,  {Simonneau} E.,  {Betancort-Rijo} J.,  {Patiri}
  S.,  {Gottl{\"o}ber} S.,    {Sanchez-Conde} M.~A.,  2006, ApJ, 645, 1001

\bibitem[\protect\citeauthoryear{{Springel} \& {White}}{{Springel} \&
  {White}}{1999}]{1999MNRAS.307..162S}
{Springel} V.,  {White} S.~D.~M.,  1999, MNRAS, 307, 162

\bibitem[\protect\citeauthoryear{{Stadel}}{{Stadel}}{2001}]{2001PhDT........21%
S}
{Stadel} J.~G.,  2001, PhD Thesis, University of Washington

\bibitem[\protect\citeauthoryear{{Tremaine}, {Richstone}, {Byun}, {Dressler},
  {Faber}, {Grillmair}, {Kormendy} \& {Lauer}}{{Tremaine}
  et~al.}{1994}]{1994AJ....107..634T}
{Tremaine} S.,  {Richstone} D.~O.,  {Byun} Y.-I.,  {Dressler} A.,  {Faber}
  S.~M.,  {Grillmair} C.,  {Kormendy} J.,    {Lauer} T.~R.,  1994, AJ, 107, 634

\bibitem[\protect\citeauthoryear{{von Neumann}}{{von
  Neumann}}{1951}]{1951NBSAM..12...36v}
{von Neumann} J.,  1951, NBS App. Math. Series, 12, 36

\bibitem[\protect\citeauthoryear{{Zemp}}{{Zemp}}{2003}]{2003DiplT.........Z}
{Zemp} M.,  2003, Diploma Thesis, ETH Zurich

\bibitem[\protect\citeauthoryear{{Zemp}, {Stadel}, {Moore} \& {Carollo}}{{Zemp}
  et~al.}{2007}]{2007MNRAS.376..273Z}
{Zemp} M.,  {Stadel} J.,  {Moore} B.,    {Carollo} C.~M.,  2007, MNRAS, 376,
  273

\bibitem[\protect\citeauthoryear{{Zhao}}{{Zhao}}{1996}]{1996MNRAS.278..488Z}
{Zhao} H.,  1996, MNRAS, 278, 488

\end{thebibliography}

\label{lastpage}

\end{document}